\documentclass[manuscript]{acmart}
\AtBeginDocument{%
  \providecommand\BibTeX{{%
    \normalfont B\kern-0.5em{\scshape i\kern-0.25em b}\kern-0.8em\TeX}}}

\setcopyright{none}

\usepackage{pbox}
\usepackage{array}
\usepackage{xcolor}
\usepackage{soul}
\usepackage{csquotes}
\usepackage{longtable}

\usepackage[normalem]{ulem}               
\newcommand\redout{\bgroup\markoverwith
{\textcolor{red}{\rule[0.5ex]{2pt}{0.8pt}}}\ULon}

\begin{document}

\title{Characterizing Barriers and Technology Needs in the Kitchen for Blind and Low Vision People}

\author{Ru Wang}
\email{ru.wang@wisc.edu}
\affiliation{%
 \institution{University of Wisconsin--Madison}
 \city{Madison}
 \state{Wisconsin}
 \country{USA}
}

\author{Nihan Zhou}
\affiliation{%
 \institution{University of Wisconsin--Madison}
 \city{Madison}
 \country{USA}}
\email{nzhou37@wisc.edu}

\email{nzhou37@wisc.edu}

\author{Tam Nguyen}
\affiliation{%
 \institution{University of Wisconsin--Madison}
 \city{Madison}
 \country{USA}}
\email{nguyen58@wisc.edu}

\author{Sanbrita Mondal}
\affiliation{%
 \institution{University of Wisconsin--Madison}
 \city{Madison}
 \country{USA}
}
\email{smondal4@wisc.edu}

\author{Bilge Mutlu}
\affiliation{%
\institution{University of Wisconsin--Madison}
\city{Madison}
\state{Wisconsin}
\country{USA}}
\email{bilge@cs.wisc.edu}

\author{Yuhang Zhao}
\affiliation{%
 \institution{University of Wisconsin--Madison}
 \city{Madison}
 \state{Wisconsin}
 \country{USA}}
\email{yuhang.zhao@cs.wisc.edu}


\settopmatter{printacmref=false}

\renewcommand{\shortauthors}{Wang et al.}
\renewcommand{\shorttitle}{Technology needs in the kitchen}
\newcommand{\TODO}[1]{\textcolor{red}{\textbf{[#1]}}}
\newcolumntype{P}[1]{>{\centering\arraybackslash}p{#1}}

\let\svthefootnote\thefootnote
\newcommand\freefootnote[1]{%
  \let\thefootnote\relax%
  \footnotetext{#1}%
  \let\thefootnote\svthefootnote%
}

\begin{abstract}

Cooking is a vital yet challenging activity for people with visual impairments (PVI). It involves tasks that can be dangerous or difficult without vision, such as handling a knife or adding a suitable amount of salt. A better understanding of these challenges can inform the design of technologies that mitigate safety hazards and improve the quality of the lives of PVI. Furthermore, there is a need to understand the effects of different visual abilities, including low vision and blindness, and the role of rehabilitation training where PVI learn cooking skills and assistive technologies. In this paper, we aim to comprehensively characterize PVI's challenges, strategies, and needs in the kitchen from the perspectives of both PVI and rehabilitation professionals. Through a contextual inquiry study, we observed 10 PVI, including six low vision and four blind participants, when they cooked dishes of their choices in their own kitchens. We then interviewed six rehabilitation professionals to explore their training strategies and technology recommendations. Our findings revealed the differences between low vision and blind people during cooking as well as the gaps between training and reality. We suggest improvements for rehabilitation training and distill design considerations for future assistive technology in the kitchen.


\end{abstract}





\maketitle








\section{Introduction}
Visual impairment is a prevalent disability that significantly affects people's quality of life. People with visual impairments (PVI) can experience different levels of vision loss, including mild low vision, moderate low vision, severe low vision, and blindness. Low vision refers to visual impairments that fall short of blindness but cannot be fully corrected by eyeglasses or contact lenses \cite{massof2001issues, corn2010foundations}. Low vision people experience different visual conditions, such as central vision loss, peripheral vision loss, blind spots, and extreme light sensitivity. In 2020, it was estimated that 255 million people worldwide had moderate to severe low vision and that 49 million people were blind \cite{bourne2020global}. 

Cooking is a vital but challenging activity for PVI, as it involves various vision-related tasks, such as locating items, pouring, measuring ingredients, prepping, cleaning, and interacting with kitchen appliances \cite{bilyk2009food,kostyra2017food}. Some cooking tasks can even pose safety risks, for example, handling sharp tools and hot appliances \cite{kashyap2020behaviors}. 
To simplify cooking, many PVI prepare frozen or pre-processed food with rich calories or eliminate cooking entirely by more frequently eating out at restaurants. As a result, PVI tend to have a poor nutritional status \cite{jones2019analysis, montero2005nutritional} or be overweight due to higher-than-average restaurant use \cite{bilyk2009food}.

Various cooking tools are available to assist PVI in the kitchen, such as liquid level indicators that warn the user when water reaches a specific level, talking thermometers that verbally report food temperature, and pot minders that ring when water boils \cite{ToolsforCooking}. However, the assistance provided by these tools is generally limited and may not fulfill PVI's full range of needs in the dynamic and busy cooking process. 
Many complex and even dangerous kitchen tasks remain unaddressed. For example, how can we enable PVI to cut food into uniform pieces efficiently? How can they use hot stove burners safely? Thus, there is a need for new technologies to improve PVI's cooking experience. 




Rehabilitation training is the primary channel where PVI learn about assistive technologies and skills for activities of daily living (ADL) \cite{markowitz2006occupational}. Many organizations for blind and low-vision people as well as rehabilitation clinics offer occupational therapy services to train PVI in a kitchen environment. K-12 schools also offer individualized education plans (IEP) for students with disabilities, which provide specialized instructions and services to help students learn necessary living skills, including those that are necessary for the kitchen \cite{IEPGuide}. 
Prior research has shown that rehabilitation training can effectively improve PVI's performance in daily activities \cite{liu2013occupational}. 
Since such training can greatly shape PVI's cooking strategies and technology choices, taking into account the views of rehabilitation professionals can enable the design of more practical and widely adopted technology for PVI. 


To inform the technology design for PVI in the kitchen, we seek to draw insights from both the PVI and the rehabilitation professionals. While some recent work has explored PVI's cooking experiences and challenges via interviews \cite{kashyap2020behaviors, li2021non} and cooking video observations \cite{li2021non}, these studies only focused on the perspective of PVI and on people who are blind. 
Furthermore, very little is known about the rehabilitation training process and the perspectives of rehabilitation professionals. For example, what cooking tasks do these professionals focus on during rehabilitation training? What technologies and strategies do they recommend to PVI? And how do they determine what technologies to recommend? Answers to these questions can significantly affect PVI's cooking experiences and technology adoption.





To fill this gap, our work triangulated PVI's challenges and needs in the kitchen by investigating their individual cooking experiences as well as the rehabilitation training process. We conducted two studies: (1) a contextual-inquiry study with 10 PVI (six low vision, four blind) to explore both low-vision and blind people's first-hand cooking experiences and challenges; and (2) an interview study with six rehabilitation professionals as well as the observation of several rehabilitation training sessions to investigate the training process and the professionals' technology recommendations. 

Our studies detailed PVI's cooking challenges, strategies, and rehabilitation training experiences for cooking, and further 
revealed the key differences between blind and low-vision people in kitchen activities. 
Given PVI's different abilities and needs, we also found that most rehabilitation professionals customized their training plans and technology recommendations to better serve their clients. However, the professionals also pointed out that the current training process had limitations and thus cannot fully address all challenges faced by PVI in the kitchen. By combining the two studies, we associated PVI's cooking experiences and challenges with the professionals' training strategies, thus identifying both technological and training gaps in the essential cooking activities.
Based on the suggestions from PVI and the professionals, we derive design implications for more effective assistive technology and more feasible training procedure to empower PVI in the kitchen.



%
\section{Related Work}
 Researchers have put significant effort to assist PVI in various daily activities, such as reading \cite{zhao2015foresee, huang2019augmented}, shopping \cite{bigham2010vizwiz, lanigan2006trinetra, zhao2016cuesee}, navigation \cite{giudice2008blind, zhao2020effectiveness, sato2017navcog3, ahmetovic2016navcog}, and socializing \cite{kramer2010smartphone, krishna2010vibroglove, zhao2018face}. However, little attention has been paid to cooking, an important but challenging activity for basic living. In this section, we summarize the existing research for PVI that is relevant to cooking.

\subsection{Food Experience of People with Visual Impairments.}
Cooking and food preparation has been shown to be one major challenge that PVI face in their daily life \cite{taylor2016does, salminen2014young, cimarolli2012challenges}. 
Salminen and Karhula \cite{salminen2014young} interviewed 14 young PVI and 22 parents to investigate their challenges and found that meal preparation was one of the major challenges and many participants felt scared of dealing with hot kitchen elements. Taylor et al. \cite{taylor2016does} reviewed 123 studies with people with age-related macular degeneration (AMD) and showed that AMD negatively affected various daily activities, such as mobility, meal preparation, shopping, and cleaning. 

Some researchers have specifically focused on PVI's cooking and food preparation experience. Bilyk et al. \cite{bilyk2009food} interviewed nine blind or severely visually impaired participants in Canada about their food experience, finding that PVI faced all kinds of obstacles when shopping for, preparing, and cooking food. They also found that the barriers in food preparation may affect PVI's health condition and lead to obesity due to higher-than-average restaurant use. 
Jones et al. \cite{jones2019analysis} conducted a 37-question survey with 101 PVI in the U.K., asking about their experiences with shopping and cooking. 
The study revealed that vision loss made cooking difficult for PVI and the difficulty level was correlated with the level of participants' visual impairments. 
Beyond cooking challenges and experiences, Kostyra et al. \cite{kostyra2017food} also explored PVI's desired technology to facilitate cooking and food preparation. They surveyed 250 PVI, including 79.6\% participants who had little to no vision and 20.4\% who could read large print, and found that PVI preferred electronic devices, such as devices that supported voice input. Prior research mostly used the method of survey and interview to understand PVI's cooking experience. Thorough observation is needed to deeply understand their cooking process.  

Recently, Li et al. \cite{li2021non} explored PVI's cooking experience via both interviews and analysis of 122 YouTube videos of PVI preparing meals. This research focused on participants who had little to no vision. As a result, they summarized 12 essential cooking activities conducted by PVI from the videos and identified several cooking-related challenges faced by PVI, such as utilizing cooking tools, accessing information in the kitchen. 
Building upon Li et al.'s research, the first study of our research adopted a different research approach---observing PVI's first-hand cooking experiences via contextual inquiry \cite{karen2017contextual}---to triangulate their cooking challenges and needs. Moreover, our study involved both blind and low vision participants, allowing us to discover the unique needs of people with different visual abilities.

\subsection{Technologies to Facilitate Cooking for People with Visual Impairments.}
PVI used various conventional tools to deal with small cooking tasks, ranging from low-tech aids to mid-tech electronic devices \cite{hersh2008assistive, ToolsforCooking}. Examples of low-tech aids included tactile or braille labels to mark kitchen elements, oven mitts to prevent burns, and pot minders that can be put into water and ring when water boils. There have also been mid-tech electronic devices that provide audio information for PVI. For example, liquid level indicators can be hung at the lip of a container and generate alert sounds when the liquid level approaches the top of the container \cite{LiquidLevelIndicator}; talking cooking thermometers can be inserted into the food and verbally report the food temperature \cite{TalkingThemometer}. Moreover, Hersh and Johnson summarized the assistive devices designed for PVI to support food preparation and consumption~\cite{hersh2008assistive} and mentioned several other mid-tech devices, such as talking measuring jugs, talking microwaves, talking scales, and a grill alert communicating remote meat thermometer.
However, conventional cooking tools only provided limited assistance for specific small kitchen tasks and cannot fully support the complex and busy cooking process. 

In the research field, limited effort has been made to facilitate cooking for PVI. 
Kim et al. \cite{kim2022vision} proposed a vision-based system that can detect burners, cookwares and the user's hands using deep learning, based on which, the system can guide the PVI to locate burners safely through auditory feedback and visual augmentations projected on the stovetop. Ramil Brick et al. \cite{ramil2021allergic} developed a voice assistant for PVI that can answer users' questions about information on food packages in natural language, such as detecting allergens and expiration date. Although the system outperformed other Visual Question Answering (VQA) systems on food packaging questions, the overall error rate was still relatively high.
Other relevant work focused on enabling PVI to interact with the digital interfaces, which can be applied to kitchen appliances \cite{morris2006clearspeech, fusco2014using, tekin2011real, guo2016vizlens, guo2019statelens}. For example, Fusco et al. \cite{fusco2014using} designed a ``Display Reader'' smartphone application that recognized the text on digital screens and read the information to PVI. The system also provided audio guidance to enable PVI to aim their phone camera properly and take usable images for recognition. With the system VizLens \cite{guo2016vizlens}, Guo et al. provided more interactive feedback by recognizing and describing the part of the appliance interface beneath the user's finger. Besides audio feedback, Guo et al. also supported digital display interaction for PVI via 3D printed tactile overlays \cite{guo2017facade}. 
For low vision people, Lang and Machulla \cite{lang2021pressing} built a smart-glass system that facilitated touchscreen interaction by generating visual guidance and highlights to attract users' attention to a specific button on the touchscreen.

Since none of these technologies have been deployed to the market, it is unclear whether and how PVI would want to use such technology in real-life cooking activities, and whether the rehabilitation professionals would recommend such technology.

\subsection{Rehabilitation Training for People with Visual Impairments.}
Vision rehabilitation refers to a range of professional training services for PVI to help them live independently and maintain their accustomed quality of life \cite{VisionRehabilitation, meyniel2017revisiting}. It is an important source for PVI to learn living skills and assistive technologies for various daily activities, such as reading \cite{palmer2010effective, seiple2011reading, smallfield2020occupational}, socializing \cite{nastasi2020occupational}, using technologies \cite{kaldenberg2017training}, and cooking \cite{hapeman2008recipe, meyniel2017revisiting}. Many blindness and low vision organizations and clinics offer such services via occupational therapy. Schools also provide the individualized education plan (IEP) to better meet the goals of students with disabilities, which includes rehabilitation training that improves students' daily living skills such as cooking \cite{goodman1993individualized}.   

Some researchers have explored the process and effectiveness of rehabilitation interventions \cite{markowitz2006occupational, goldstein2015clinically, liu2013occupational, kaldenberg2020occupational}. For example, Liu et al. \cite{liu2013occupational} reviewed 17 empirical studies on occupational therapy training to explore its impact on the activities of daily living (ADL) for older adults with low vision. They found that multi-component and multi-disciplinary interventions for personal goals had more positive outcomes than non-personalized training. Keldenberg and Smallfield \cite{kaldenberg2020occupational} reviewed 38 papers on different intervention approaches, showing that multi-component rehabilitation can facilitate ADL performance. 
Moreover, Smallfield and Kaldenberg \cite{smallfield2020occupationalcase} reported a clinical case of low vision occupational therapy with a patient who experienced dry age-related macular degeneration, revealing the detailed evaluation and training process in the intervention.
However, none of the aforementioned research focused on the training for cooking activities.

Little research has investigated rehabilitation training for cooking. Miller \cite{miller2001use} explored the outcome of seven teenagers who attended a summer rehabilitation program at the Vacation Camp for the Blind. The study showed that two participants received training for cooking and both reported improvements in their kitchen skills, such as the safety of preparing hot meals and using a stove. Freyberger \cite{freyberger1971comparative} highlighted the differences between congenitally blind people and people who acquired blindness later in their life and suggested different training methods for them during cooking. For example, the training approaches for congenitally blind people should be basic, while people with acquired blindness only need to relearn known cooking skills. In 2005 and 2006, Milwaukee's Solomon Juneau Business Charter High School collaborated with Badge Association of the Blind and Visual Impaired to offer cooking instruction to students with visual impairments as a supplementary program to the general curriculum \cite{hapeman2008recipe}. The program was a success, demonstrating the progress and confidence gained by students with visual impairments. 
 
 Unlike prior research that investigated the training approaches from the point of view of vision science, we explore the rehabilitation training process for cooking from the lens of Human-Computer Interaction (HCI). By interviewing rehabilitation professionals and observing their training sessions, we seek to understand their training principles, the decision-making process for training plans, technology recommendations, and their vision on future technologies. By combining PVI's first-hand experiences with the professionals' feedback, we derive design considerations to inspire technologies that enable PVI to cook safely and independently in the kitchen.  

\section{Study 1: Contextual Inquiry Study with People with Visual Impairments}
We first explored PVI's cooking experience via first-hand observation. We conducted a contextual inquiry study with 10 participants with visual impairments by observing their cooking process in their own kitchen, thus understanding their challenges, strategies, and technology needs. 

\subsection{Method}

\subsubsection{Participants.}
We recruited 10 participants with visual impairments (3 male, 7 female) whose ages ranged from 18 to 73 ($mean = 54.1, SD = 18.3$). Four participants were completely blind, while the other six had low vision (LV). Except for Dixie, all other participants were legally blind, meaning that their visual acuity was 20/200 or less in the better eye with best correction or their field of view was less than 20 degrees \cite{LegalBlind}. Six participants had congenital visual impairments (i.e., visual impairment at birth), and four experienced acquired visual impairments (i.e., vision loss that happened to people who were originally sighted \cite{AcquiredVI}). Table \ref{tab:pvi_dem} shows participants' demographic information.

We supported both in-person and remote studies due to the constraints caused by the COVID-19 pandemic. We recruited local participants by distributing recruitment flyers at the local low vision services at the Department of Ophthalmology and Vision Sciences, the University of Wisconsin-Madison. We also recruited remote participants by sending out recruitment emails via the National Foundation of the Blind (NFB)~\cite{NFB} and posting recruitment information on various online communities, such as Reddit /rblind and the Top Tech Tidbits newsletter~\cite{tttnews}. A participant was eligible for our study if they had visual impairments and had cooking experience. 



\begin{table}[ht]
\footnotesize
\centering
\caption{Demographic information of the 10 participants in Study 1.}
\begin{tabular}{C{0.8cm}C{0.5cm}C{0.7cm} C{1.8cm} C{1.6cm} C{1.3cm} C{1.2cm} C{2.4cm} C{1.6cm}}
\toprule
\textbf{Pseudonym}&\textbf{Age/{\newline}Sex}&\textbf{Blind  or LV}&\textbf{Diagnosed{\newline}Condition}&\textbf{Visual{\newline}Acuity}&\textbf{Visual{\newline}Field}&\textbf{Onset{\newline}Age}&\textbf{Cooking {\newline}Training}&\textbf{Demonstrated \newline Dish}\\
\midrule
Hilary&24/F&  LV & Albinism &Left: 20/500 \newline Right: very low & Peripheral vision loss &Congenital&Occupational Therapy&Vegetable Tofu Stir Fry\\ 
\hline
Chloe & 18/F& LV & Autoimmune Chorioretinitis &  Left: 20/250 \newline Right: 20/300 & Central vision loss & 16 & Occupational Therapy & Enchiladas \\ 
\hline
Dixie & 73/F & LV & Hemianopsia & 20/20 to 20/40 & Peripheral vision loss  & 70 & Occupational Therapy {\newline}(only on mobility) & Chicken Soba Noodle\\ 
\hline
Derek & 60/M & LV & Diabetic Retinopathy, Glaucoma  & Motion and light perception only & Peripheral vision loss & 50 & Occupational Therapy & Microwave Frozen Food\\ 
\hline
Judy & 59/F & LV & Spinal Meningitis  & Left: 20/2200 \newline Right: 20/400 & Peripheral vision loss & Congenital & N/A & Lasagna \\ 
\hline
Jacob & 55/M& LV & Anterior Ischemic Optic Neuropathy  & Left: 20/40 \newline Right: 20/200 & Peripheral vision loss & 52 & Blind Camp & Breakfast Burrito\\ 
\hline
Kim & 59/F& Blind & Leber's Congenital Amaurosis  & - & - 
& Congenital & N/A & Banana Pancake\\ 
\hline
David & 59/M& Blind & - & - & - & Congenital & Adjustment to vision loss program at college & Pizza \\ 
\hline
Rene & 67/F& Blind & -  & - & - & Congenital & N/A & Lasagna\\ 
\hline
Mary & 67/F& Blind& Congenital Cataract    & - & - 
& Congenital & College classes for TVI training; Blindness adjustment training & American Breakfast\\ 
\bottomrule
\end{tabular}

\label{tab:pvi_dem}
  \vspace{-3ex}

\end{table}

\subsubsection{Procedure.}
Our study consisted of a single session that lasted two hours. The study included three phases: an initial semi-structured interview, a cooking observation, and a post-observation interview. 

In the initial interview, we asked about participants' demographic information (e.g., age, gender), visual conditions, and use of assistive technologies. We also asked about participants' general cooking experience and habits, such as how often they cook, whether they cook independently, whether they have cooked raw food or only processed food, and what kinds and styles of food they usually cook. Participants also talked about their prior rehabilitation training experience for cooking. 


In the observation phase, we observed participants cooking a meal of their choice following the master-apprentice model of contextual inquiry \cite{karen2017contextual}. Participants first showed us around their kitchen and explained the customization they made to the kitchen if there was any. We then asked them to start cooking and observed their behaviors during the whole process, from food preparation to cooking to cleaning.  
Participants were asked to ``think aloud'' when they carried out different tasks in the kitchen, describing what they were doing, what tools and strategies they were using, what challenges they were facing, and why. 
When we noticed something interesting, we also conversed with the participants to understand their intentions, behaviors, and the rationales behind their behaviors.  

In the post-observation interview, 
We first asked about participants' experiences with other common tasks in the kitchen that were absent from the observation. 
We compiled a list of common cooking tasks based on existing cooking guidelines and practices for PVI \cite{SafeCooking, tzerniadakis2016research} and PVI's cooking challenges \cite{li2021non, bilyk2009food, kashyap2020behaviors, jones2019analysis}. Based on this list, we identified the tasks that participants didn't demonstrate in the observation session and asked participants to describe their experiences with these tasks, including their challenges and coping strategies. The list is shown in Appendix Table \ref{tab:tasks}.

After discussing each cooking task, we asked participants about the hardest and most time-consuming task they encountered in the kitchen. Participants then rated their general cooking experience on different aspects via 5-point Likert-scale scores, including both physical (i.e., physical tiredness, safety) and psychological status (i.e., satisfaction, confidence, mental stress). Finally, we ended our study by asking about participants' suggestions on current kitchen tools as well as their desire for future assistive technologies in the kitchen. 

\subsubsection{Study Setup.}
We offered participants both in-person and remote study options given the pandemic. If the participant chose an in-person study, we visited the participant's home and observed their cooking process in their own kitchen. In terms of remote study, we set up a Zoom video call by asking participants to place their cameras (e.g., cameras on their phone or laptop) in their kitchen. We also communicated with the participants via Zoom before the formal study to help them adjust the position and angle of the camera, ensuring that the camera clearly capture their cooking space and behaviors (e.g., place the camera near the stovetop). 
As a result, five studies were conducted in person, while the other five were remote.

\subsection{Analysis.}
Researchers took notes and video recorded the whole study with Zoom (remote) or a camcorder (in-person). We used an online automatic transcription service to transcribe the videos and then cleaned the data by manually correcting the auto-transcription errors. Researchers also watched the videos and annotated participants' behaviors and relevant kitchen objects in the transcripts.  

We analyzed the transcripts using the qualitative analysis method \cite{saldana2021coding, braun2006using}. We first developed codes using open coding. Two researchers independently read three identical transcripts and watched the corresponding videos to code the data. After the two researchers compared and discussed the codes, a codebook was generated based on their agreement. Each researcher then coded half of the remaining transcripts based on the agreed codes. If new codes emerged, the two researchers discussed again and updated the codebook upon agreement. The codes were then categorized using axial coding and affinity diagram. Our analysis resulted in 12 themes (Table \ref{tab:PVI_themes}).  

We also quantitatively analyzed participants' Likert scale scores on their cooking experiences. To compare the cooking experiences between participants with different visual conditions, we define two between subject factor: \textit{\textbf{VisionLossOnset}} (\textit{Congenital} vs. \textit{Acquired}) and \textit{\textbf{VisualAbility}} (\textit{Blind} vs. \textit{LowVision}). We analyzed the effect of VisionLossOnset and VisualAbility respectively on the six sets of Likert scale scores, including physical tiredness, safety, mental stress, satisfaction with food, satisfaction with technology, and confidence (Figure \ref{fig:avicvi_likert}-\ref{fig:blv_likert}). We used the Wilcoxon rank sum test since the distributions of the scores were not normal. We also calculated the effect size $r$, with the thresholds of small, medium and large effects being 0.1, 0.3, 0.5, respectively~\cite{cohen2016power, cohen2013statistical}.




\subsection{Findings}
\subsubsection{\textbf{Cooking Experiences of People with Different Visual Conditions.}}
All participants were able to cook some dishes independently but their cooking abilities varied. Six participants indicated having more than 40 years of cooking experience. 
Eight out of 10 participants had cooked raw and fresh food.
Derek and Chloe did not have much cooking experience but they could cook pre-processed and frozen food using a microwave. 
Participants had cooked various types of food: five participants (Judy, Derek, David, Rene, and Mary) mainly cooked American food, while others cooked Mexican food (Jacob), Asian food (Hilary), and mixed style of food (Dixie, Kim, and Chloe).


  

\begin{figure}
  \begin{minipage}{0.5\textwidth}
        \centering
        \includegraphics[width=0.92\textwidth]{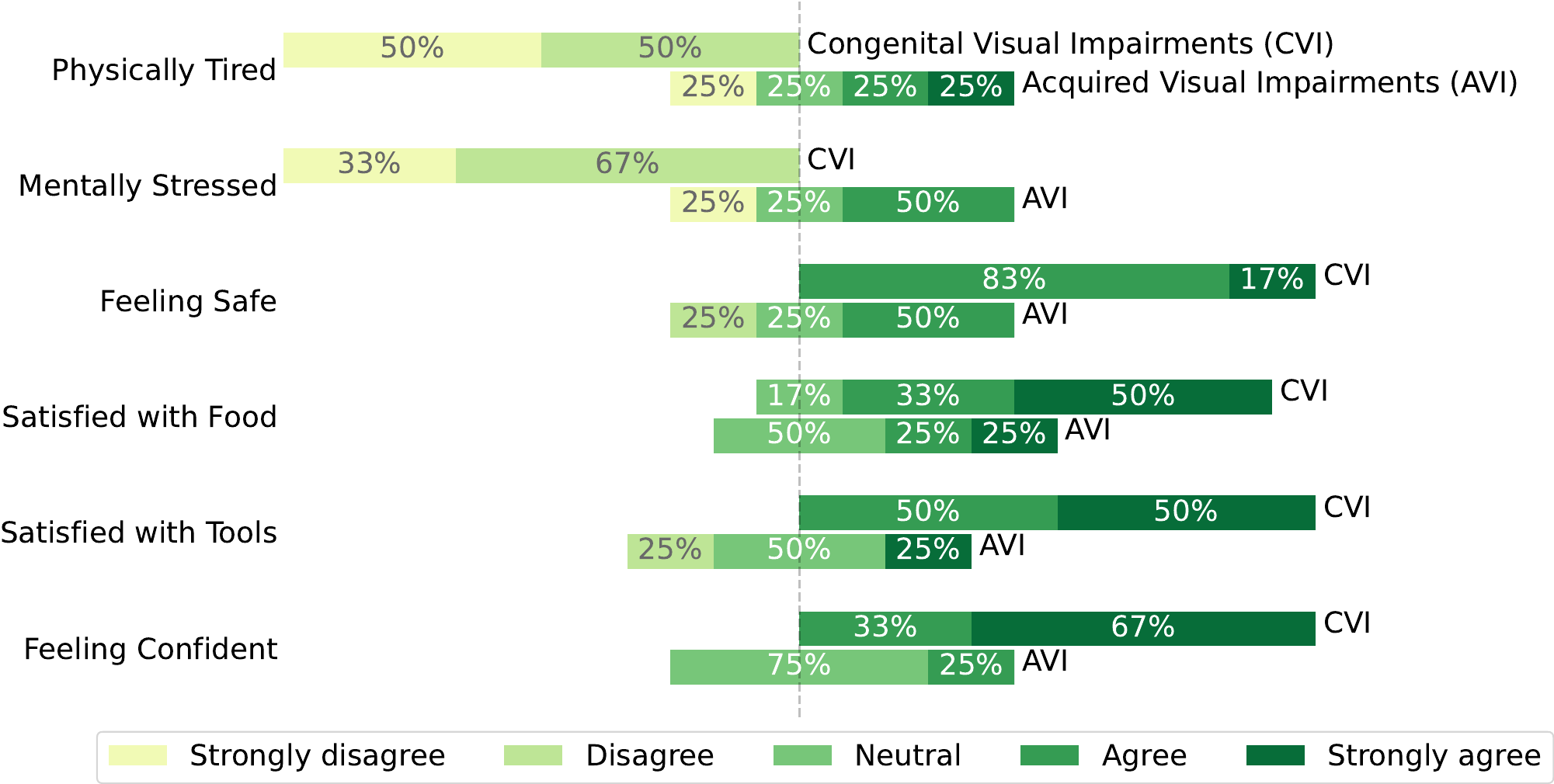} 
        \caption{5-point Likert-scale self-assessment on the cooking experience between participants with CVI and participants with AVI}
        \label{fig:avicvi_likert}
    \end{minipage}\hfill
    \begin{minipage}{0.5\textwidth}
        \centering
        \includegraphics[width=0.92\textwidth]{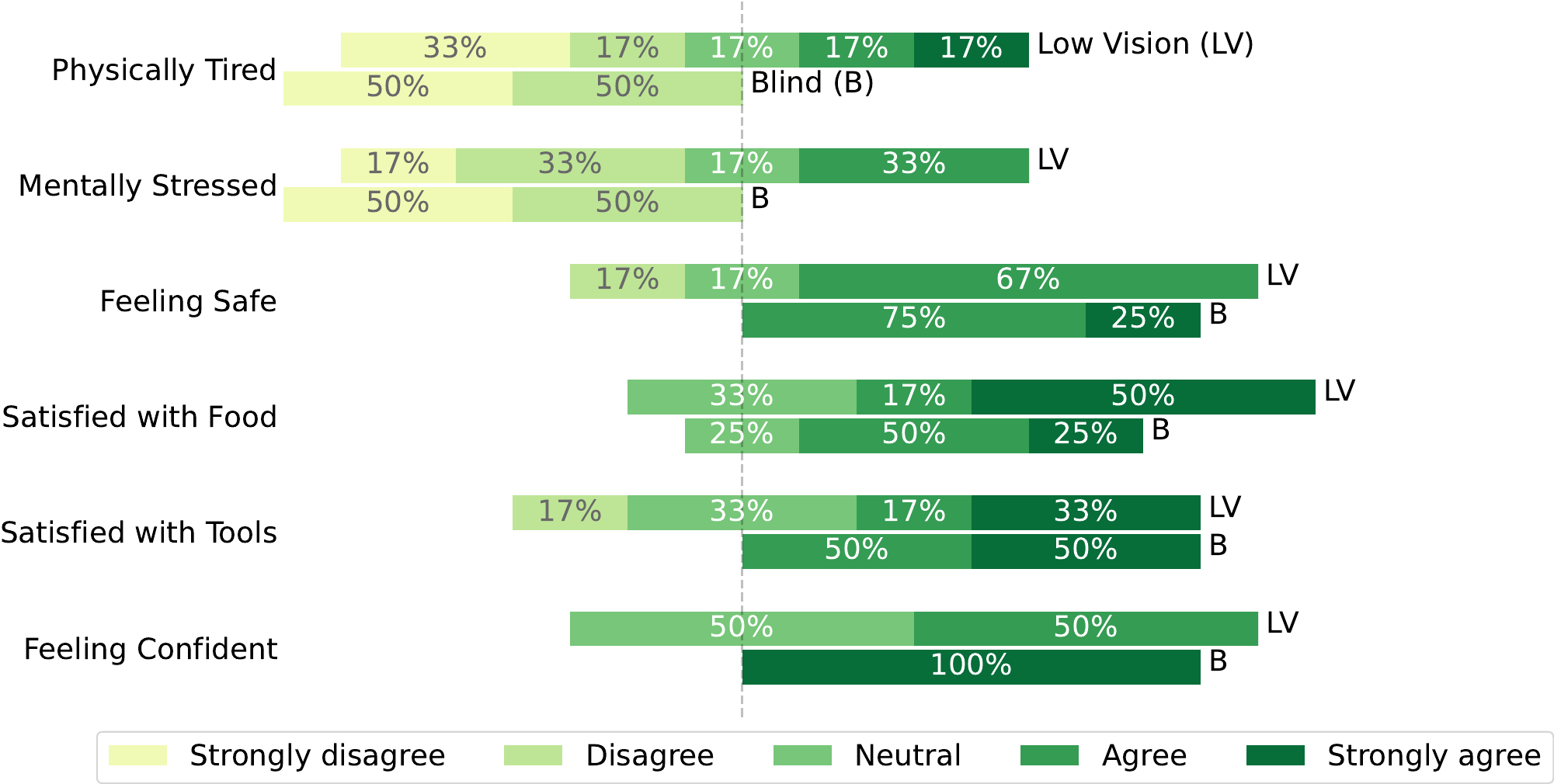} 
        \caption{5-point Likert-scale self-assessment on the cooking experience between blind and low vision participants}
        \label{fig:blv_likert}
    \end{minipage}
    \vspace{-3ex}
\end{figure}

We found that participants' cooking experiences differed due to their visual conditions, including the onset time of visual impairments (congenital vs. acquired visual impairments) and their residual vision (blind vs. low vision). We compared the self-assessment scores on the cooking experience between participants with different visual impairment onset time and residual vision, as shown in Figure \ref{fig:avicvi_likert} and \ref{fig:blv_likert}, respectively.
 Compared to participants with congenital visual impairments (CVI), participants with acquired visual impairments (AVI) felt significantly less confident about their cooking skills (AVI: $mean=3.25, SD=0.50$; CVI: $mean=4.67, SD=0.52$; Wilcoxon rank sum test: $W = 24$, $p = 0.01$, $r = 0.86$). While no significant difference was found, participants with AVI indicated more safety concerns than those with CVI (AVI: $mean=3.25, SD=0.96$; CVI: $mean=4.17, SD=0.41$).
Several participants with AVI (e.g., Dixie, Chloe) even expressed fear of injuries during cooking due to the progression of their vision loss. 
As Dixie described, 
\textit{``The food blender blade scares me, and I have a coffee grinder that I’m afraid of... My vision [loss] has enhances the fear.''}
 
 Moreover, since people with AVI experienced vision loss later in their life, they had to re-adapt their living skills, which resulted in a trend of higher mental stress (AVI: $mean=3.00, SD=1.41$, CVI: $mean=1.67, SD=0.52$)
and physical tiredness (AVI: $mean=3.25, SD=1.71$, CVI: $mean=1.50, SD=0.55$), as well as less satisfaction with their food (AVI: $mean=3.75, SD=0.96$, CVI: $mean=4.33, SD=0.82$) and tools for cooking (AVI: $mean=3.25, SD=1.26$, CVI: $mean=4.50, SD=0.55$); however, no statistical significance was found.
 Specifically, four participants with AVI (Chloe, Derek, Dixie, and Jacob) gave high scores ($\geq 3$) to their stress and tiredness level. They felt frustrated about many small tasks in the kitchen and the consequences of failing the tasks. For example, 
Derek expressed the frustration of dropping a fork on the floor and the stress just by thinking about searching for it: \textit{``I'm so frustrated about that. And sometimes they bounce in big travel and I don't know where the heck it goes.
 And I can't leave things on the floor because I don't want to trip. I don't want it to mess me up.''}
 
Since most low vision participants experienced acquired vision loss, we found that low vision participants were significantly less confident than blind participants (LV: $mean=3.50, SD=0.55$, Blind: $mean=5.00, SD=0$; Wilcoxon rank sum test: $W = 24$, $p = 0.01$, $r = 0.86$). It was worth noting that, all blind participants gave a high score of five, indicating their confidence. Moreover, while low vision participants and blind participants reported similar satisfaction towards their food (LV: $mean=4.17, SD=0.98$, B: $mean=4.00, SD=0.82$), we found a trend that low vision people felt less safe (LV: $mean=3.50, SD=0.84$, B: $mean=4.25, SD=0.50$), more physically tired (LV: $mean=2.67, SD=1.63$, B: $mean=1.50, SD=0.58$), and more mentally stressed (LV: $mean=2.67, SD=1.21$, Blind: $mean=1.50, SD=0.58$) during cooking.  
Importantly, low vision people also showed a trend of being less satisfied with the current cooking tools than blind people (LV: $mean=3.67, SD=1.21$, Blind: $mean=4.50, SD=0.58$), which indicated a need for cooking technology that considers low vision people's unique needs. However, no statistical significance was found except for the confidence level. 

\subsubsection{\textbf{Learning Cooking Skills.}}
Learning how to cook was challenging to PVI due to the difficulty of learning by observing. As David explained, 
\begin{displayquote}
\textit{``It's interesting... Because I work in education, every sighted person learns so much just from observing, even things like social, like looking at somebody when they are talking or knowing what distance to be from somebody to not getting in their personal space. All of that needs to be kind of deliberately taught to blind students because they do not learn so much from observation. I'm a transition specialist, so I work with students 14 up to prepare for life after high school... A sighted kid can go into McDonald's and see people behind the counter cooking and see people at the cash register and somebody putting food in the bag. But blind children have to be taught so much more.''}
\end{displayquote}

As such, an important approach to learn cooking was through rehabilitation programs. Some participants in our study received rehabilitation training---six participants were trained specifically for cooking, and one (Dixie) only participated in orientation and mobility (O\&M) training. Participants obtained training from different services, including occupational therapy (Chloe, Hillary, Derek), blindness adjustment classes in college (David, Mary), and a blind camp run by Office for the Blind and Visually Impaired (Jacob). 

Five participants learned strategies for safe cooking during their training, such as cutting with proper hand postures and wearing gloves when dealing with heat. Some participants were also trained for other cooking tasks. For example, David learned measuring skills and Derek learned how to interact with kitchen appliances. In addition to cooking-specific skills, some participants received training for general skills that can be used in kitchen activities, including O\&M training (Jacob, Mary, Dixie) and reading (Chloe). 

Beyond fundamental skills, rehabilitation training also involved recommending and teaching assistive technology and PVI-friendly kitchen tools. Most recommended technologies were low-tech tools, such as a slicer to cut even slices, tactile dots to label kitchen elements, talking thermometers, heatproof gloves, and rack push pullers. Derek and Jacob also showed us the George Forman Grill recommended by their therapists to prevent burning,
\textit{``It's basically just a folding frying pan that both sides get hot. So instead of trying to fry a burger on the stove, cooking a hamburger in this is so, so much easier. I don't burn myself, it prevents burning, it prevents a big mess.''} Some participants learned more advanced assistive technologies, such as digital magnifier and OrCam \cite{OrCam} (Chloe). 

Compared to a one-on-one training, Jacob found it helpful to discuss and share information with other PVI in a group training, \textit{``I enjoyed working with [other] low vision patients, so I would enjoy a session and some training, particularly with people that have similar problems. I think the power of sharing challenges in the kitchen is more than anything... Sharing is valuable to everybody who's involved.''}


Unfortunately, not all participants managed to get rehabilitation training. We found that \textbf{accessing training resources was a challenge} to some participants because they did not know who to reach out to and if they were eligible for the training (e.g., Judy). 
Some participants especially emphasized the limited resources for low vision people. Dixie had participated in a 6-week training session but she did not gain any follow-up training despite her needs for further help. 

Other than rehabilitation training, many participants (except Dixie and Derek) mentioned that they learned kitchen techniques from their family or by themselves. As Mary explained, \textit{``The majority of what I've learned [is by] just doing it. 'hey, I made a mistake, how could I do it different[ly]'''}. Judy and Jacob also mentioned learning from other PVI to discover effective strategies and new technologies in the kitchen: \textit{``I know there are things (assistive tools). but I don't know what those things are, and I don't know what questions to ask...I like talking with other visually impaired people just to find out their experiences and what things they use, what apps they use.''}

\subsubsection{\textbf{General Challenges \& Strategies for Blind and Low Vision People.}}
\label{study1:challenges_strategies_blv}
Participants identified the hardest and most time-consuming cooking tasks. Some participants felt cutting (4 participants) and dealing with heat (3) to be the hardest due to safety concerns. Other difficult tasks pointed out by participants included measuring (3), looking for ingredients (3), and checking food doneness (2). As for most time-consuming tasks, seven participants (except Chloe, Kim, Mary) mentioned food preparation, especially chopping (4). 
We report the main challenges participants faced and the common strategies adopted by both blind and low vision participants during cooking; the unique strategies used by low vision people are reported in Section \ref{study1:low_vision_section}. 

\begin{figure}
  \includegraphics[width=\textwidth]{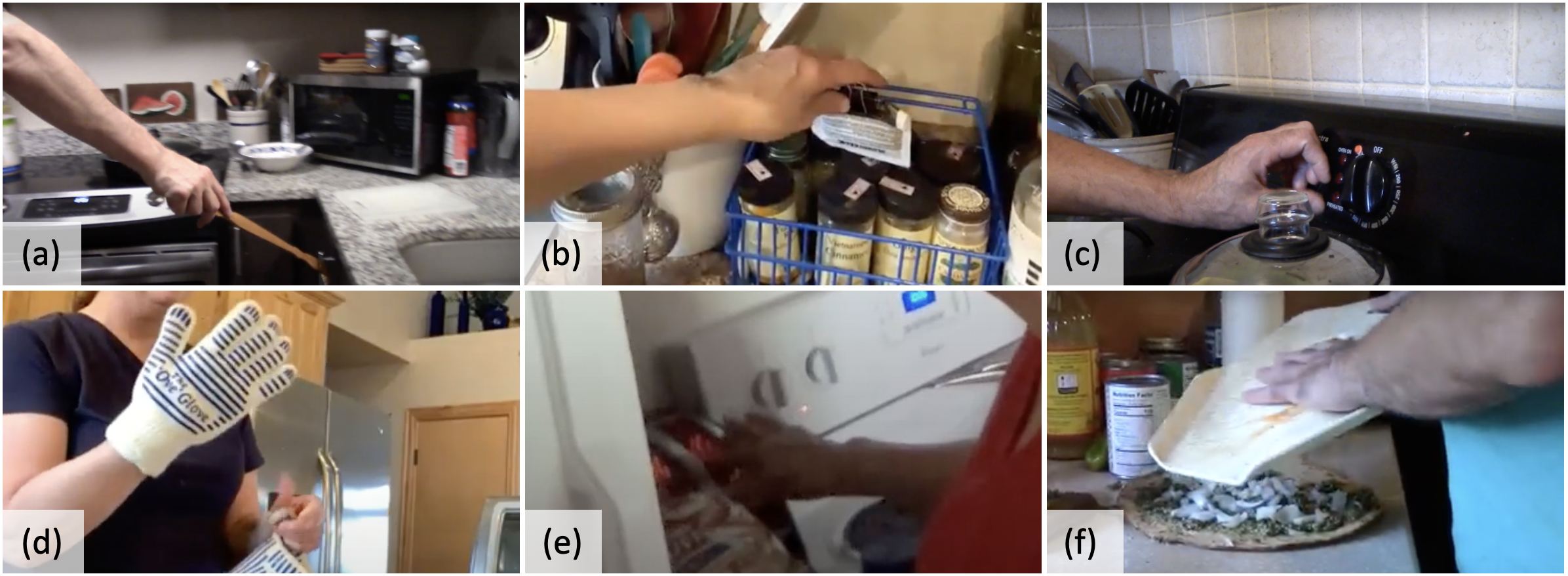}
  \caption{(a) Derek used a back scratcher to scan kitchen elements above waist; (b) Kim kept utensils in a bucket and commonly-used spices in a basket; (c) David used bump dots to label knobs; (d) Kim showed us oven gloves with fingers; (e) Mary touched the toaster slot to tell if the toast was inside; (f) David used a cutting board with chute to prevent spilling when transferring food.}
  \label{fig:general_strategy}
    \vspace{-3ex}

\end{figure}

\paragraph{\textbf{Kitchen Navigation and Exploration.}}
\label{findings:nav}
Navigating the kitchen was challenging since the kitchen was small and crowded with various tools and equipment. While many PVI were familiar with their own kitchen, participants who had recent vision loss (Derek and Dixie) found it difficult to safely navigate in the kitchen. 
For example, Derek had trouble identifying the edges of furniture. Since a long white cane was not effective to detect objects above waist level, he used a back scratcher to scan surrounding kitchen elements above the waist (e.g., an extruded drawer, Figure \ref{fig:general_strategy}a). 
The uncertainty in complex kitchen activities made navigation more difficult. 
For example, Kim mentioned that she would never walk around barefoot in the kitchen because if she dropped a glass, that would be a ``big safety consideration.''

Locating kitchen elements was another important but challenging aspect of kitchen navigation. Participants used various methods to organize their kitchen environments. Most participants (except Dixie) kept their kitchen well-organized (e.g., based on frequency or kind) to find things easily. For example, Kim kept frequently-used spices in a basket (Figure \ref{fig:general_strategy}b), while Derek always placed the lid to the right of the jar.
To distinguish and find kitchen elements, participants leveraged different sensory channels, such as feeling the shape and weight of an object (e.g., Jacob, Kim), labeling kitchen elements with braille (David, Mary), distinguishing ingredients by smell (Mary, Kim, Chloe), and shaking a container to hear the sound of food (Rene, Mary). 



In spite of the various strategies above, we still observed that participants \textbf{had difficulty adapting to the dynamic position changes of kitchen objects during the busy cooking process}. 
For instance, Mary cooked scrambled eggs with a fork. Every time she went from countertop to stovetop, she lost track of the fork she used and had to grab a new one. To better keep track of things, Derek kept frequently-used kitchen tools in his pocket while cooking, while Kim always put the tools back right after use, so that she did not need to remember where they were afterward.

\paragraph{\textbf{Dealing with Heat.}}

Dealing with heat, especially using stoves and ovens, was a big challenge for all participants. For example, 
we observed that locating a hot pan on the stove was extremely difficult and the only strategy was to touch around to find it (Mary and Kim), which led to the risk of burns. For example, Kim once burnt her hand when she tried to locate the hot pan. 
Flipping hot food in the pan was also challenging, and it was hard for PVI to tell if the food was flipped successfully. We found that Jacob and Mary who demonstrated flipping in our study both touched the hot food with their hands. 
To prevent burns, some participants (e.g., Judy) intentionally used medium to low heat to cook. 

Other strategies were also adopted to use stovetop safely. 
For example, Mary and Dixie usually turned the handle of a hot pan to the side to avoid a potential collision; Mary cleared out the stovetop to ensure that no paper was left on the burners before she turn them on. Moreover, participants preferred a gas stove over an electric stove since it cooled down faster and provided audio feedback (i.e., the sound of flame). 


The use of the oven also presented safety risks, especially putting food into and taking them out of the oven. To avoid touching hot oven racks, participants used rack push puller (e.g., Jacob, Mary) and oven mitts/pads (Judy, Rene and Mary).
 Compared to oven mitts, some participants (e.g., Kim) preferred gloves with five fingers (Figure \ref{fig:general_strategy}d) for flexibility and better sensation.
However, \textbf{the protection tools did not cover users' whole arms and participants can still get burnt.} For example, David's arm once hit the top rack in the oven and got burnt when he took food out. 

Despite the safety risks, we observed that \textbf{some participants tended to be reckless and accepted the fact that they had to sacrifice safety to cook independently}. As Rene described, 
\textit{``To see what it's doing, to see how done the meat is..., I use my finger as a reality check... And sometimes you do get burned a little bit, and sometimes you misjudge yourself, and you actually touch a hot part of the pan... If you're going to be reckless, if you're going to be doing these things, you're going to expect to get burned sometimes.''}



\paragraph{\textbf{Using Sharp Tools---Cutting, Peeling, and Shredding.}}
Many kitchen tools were sharp, which brought safety concerns to PVI. Participants reported using different sharp tools, such as knives, peelers, and graters.  

Cutting different food led to different difficulties. For example, Judy and Rene found chopping small pieces of food (e.g., garlic gloves) was time-consuming since they ``jumped around'' and were hard to collect. Meanwhile, cutting hard food (e.g., avocado pit) was more dangerous because the knife could slip. Jacob and David thus avoided cutting such ingredients. In order to cut safely, most participants (e.g., Jacob, Hilary) curled their fingers when cutting. Another strategy was to use a small knife because it was easier to control (David). 




Compared to cutting, some participants (Mary, David) \textbf{felt peeling and shredding more challenging since the fingers can get close to the blade more easily}. David shared his experience using a grater, \textit{``I just try to keep my fingers away [from the blade]. One time on this side here (he showed the side of the grater with bigger holes), I was cutting cucumbers or something. I had a bad cut. My finger went against it. And I don't really use it too much anymore.''} Some participants (e.g., Kim) used a food chopper or a blender to eradicate the risks, however, scraping food off the blade can be another risk. 

\paragraph{\textbf{Interacting with Kitchen Appliances.}} 
Five participants emphasized the difficulty of interacting with advanced kitchen appliances (e.g., oven, microwave) due to the inaccessible text label on the control components of the appliances. Many modern appliances also embedded flat buttons or touchscreens, completely removing the tactile feedback from the traditional controls (e.g, physical buttons, knobs). 

Participants used different strategies to identify buttons on appliance interfaces. Most participants (e.g., Rene, David) used bump dots or braille to mark important buttons or knobs (Figure \ref{fig:general_strategy}c). However, \textbf{the tactile labels were not always reliable}. They often fell off (David and Derek) and were hard to distinguish when multiple tactile labels were used for different buttons (Rene and David).  
For interfaces with tangible buttons, some participants (e.g., David, Kim) memorized the buttons' positions, while some (e.g., Mary) used their hands to feel around and distinguish different buttons. 
Instead of interacting with the appliance interface directly, two participants (Jacob and Derek) used a smart speaker to control the appliances with speech commands.

Besides the inaccessible input, the lack of accessible feedback also made the kitchen appliances hard to use. With all information presented visually, there was no effective way for participants to confirm the current setting and status of the appliances, such as the temperature of an oven. \textbf{Without sufficient feedback, a kitchen appliance can turn into a safety hazard}. Taking Mary using a toaster as an example: When she put the toast in the toaster and pushed down the lever, the lever was not fully pressed to the bottom and the toast was accidentally popped out without notification. After we informed her of the situation, she had to use her hands to feel around the heated toaster slot to locate the toast (Figure \ref{fig:general_strategy}e). To better perceive the appliance status, participants strived to leverage all possible audio feedback from the appliances. For example, when adjusting the oven temperature, some ovens generated an audio beep every time when a user pressed the temperature adjustment buttons (up or down). Participants (Rene, Judy, and Mary) thus tracked the temperature by counting the number of beeps. However, not all appliances provided such feedback.




Furthermore, \textbf{learning to use the appliances independently was a challenge for PVI} because the manuals were too visual (Rene and Hilary). 
Moreover, some appliances had complex interfaces with a large number of buttons, further increasing the learning curve (Derek and Mary). 
Rene explicitly expressed the need for human assistance when learning new appliances.

\paragraph{\textbf{Fine-grained Ingredient Manipulation---Transferring and Measuring.}} 
We also observed participants' difficulties with transferring and measuring ingredients.
Transferring food from one place to another was challenging because it was hard to aim at the target location precisely. 
While participants could feel the target container with one hand during transferring (Rene, Judy, Kim), it became impossible if both their hands were busy. As Rene mentioned, \textit{``When I have a pot that has two handles on the side, and I got to pour something from the pot [to] something else, or if I have something that's really hot in a pot, then I don't know where [to pour], I don't have any hands to check to see where it's going.''} 
Strategies were developed by the participants to prevent spilling when transferring food. For example, David used a cutting board with a chute to collect and transfer small food pieces (Figure \ref{fig:general_strategy}f), Chloe put the two containers side by side when transferring, 
and David and Mary transferred liquid to the target container over a much bigger container (e.g., a bowl) and re-collected the spilled liquid in the bowl using a funnel.

Participants also reported difficulties with transferring an accurate amount of ingredients---measuring. Participants (e.g., Derek) preferred multiple-size measuring spoons over a single measuring cup with ticks since the visual ticks were not accessible. To verify the desired amount of ingredients, participants (e.g., Rene, Hilary) kept their fingers at the top of the measuring spoon to feel the ingredient level. We found that liquid was more difficult to measure than solid ingredients. For solid ingredients, such as protein powder, participants (David, Hilary, and Chloe) could simply fill the measuring spoon and level off the excess. However, for liquid, such as soy sauce, participants (Rene, Kim, and Hilary) had to feel the liquid with their fingers constantly until it reached the desired level. David and Hilary also mentioned using a liquid level indicator to receive a notification when the liquid reached a certain level. 
 Due to the difficulty of measuring, Jacob, David, and Mary avoided this task as much as possible.

\paragraph{\textbf{Following Recipes.}} 
Reading and following recipes was another challenge since most recipes were presented via text and images.
Blind participants mainly relied on reading digital recipes via screen readers. David and Derek also used smart speakers (e.g., Alexa) to read recipes step by step. Moreover, David made braille recipes by himself via braille printers or embossers.
Besides the recipe presentation, we found that the instructions provided by recipes were also vision-centered (David, Jacob) with many visual descriptions, such as ``bake until golden brown''(Jacob), making it hard for PVI to follow. 

As a result, we found that most participants stuck with a set of routine recipes and did not want to experiment with new dishes. Some participants felt trying new recipes to be a low priority given the already challenging cooking process (e.g., David, Rene), and some did not feel confident to cook new dishes (e.g., Judy). However, some low vision participants mentioned trying new recipes regularly, but they usually focused on simple recipes ``that don't have a lot of steps and don't need to be super precise'' (Hilary). 


\paragraph{\textbf{Food Safety and Quality.}} 
Despite the various challenges in cooking, participants were in general satisfied with the food they cooked ($mean=4.1, SD=0.88$, Figure \ref{fig:avicvi_likert}). Some participants (Kim and Jacob) were proud of what they accomplished after cooking. However, we also found participants had low expectations on the taste of the food they cooked (e.g., Mary, Derek). As Derek commented, \textit{``I've just decided I'm not going to be too picky about things and not have the same higher expectations [as before]. I just guess to take it off the stove. And if it's too cold, I may just eat it...and if it's too hot, you can just let it cool down.''}

Food safety was a significant concern for our participants.
For example, participants had difficulty distinguishing the mold on their food. 
During our observation, we also noticed that some participants mishandled their food unconsciously, which can cause food safety issues\footnote{We intervened if noticed potential safety issues.}. For example, when unpacking the pizza crust, David was not aware of the inedible silicon gel on the crust.

Determining food doneness was important for both food safety and taste. However, six participants reported difficulty checking food doneness, including both blind (e.g., Derek, Chloe) and low vision (e.g., Jacob, Judy) participants. 
To check food doneness, participants leveraged different senses, including feeling the food with hands (Rene) or utensils (Mary, Kim), smelling (e.g., David, Hilary), listening (e.g., Rene tapped the crust of bread and listened to the sound), and tasting the food directly (e.g., Mary). Participants also adopted technologies to check food doneness by tracking the cooking time and food temperature (e.g., Chloe, David). For example, Jacob connected a smart thermostat to his smart speaker, Alexa~\cite{Alexa}, to monitor the internal temperature of food. Additionally, David used Be My Eyes~\cite{bemyeyes} to consult remote-sighted agents for food doneness. Section \ref{study1:tech_kitchen} details participants' technology use.

Several participants (Kim, Rene, Jacob) pursued food aesthetics but found it difficult to achieve. 
Participants reported the challenges of cutting uniform pieces (e.g., David, Judy) and spreading ingredients evenly on dishes (e.g., Jacob, Rene). Some participants even expressed disappointment in not being able to cook food that was aesthetically appealing. As Jacob mentioned, \textit{``It’s kind of depressing. I kind of get sad when we cook a fried egg [that looks bad]... well, especially in a restaurant, you’re proud of serving somebody an egg that just looks nice.''}

\paragraph{\textbf{Kitchen Hygiene.}} 

Participants found it frustrating to clean dishes (Mary, Rene, and Chloe) because they could not easily identify dirty spots. Four participants (e.g., Judy, Mary) touched around the dish to find the dirty spot. Some participants (e.g., Chloe, Kim) even had to over-clean the kitchen to avoid missing any dirty spots. 
Cleaning knives was more challenging due to safety concerns. To avoid being cut, participants usually searched for knives and other sharp tools in the sink and cleaned them first before soaking all the dishes.
 As David explained, \textit{``This [blender blade], you have to be careful. I always wash this one right away because I don't put it in the water, so I don't get caught on it.''}



\subsubsection{\textbf{Unique Experiences of Low Vision People in the Kitchen.}}
\label{study1:low_vision_section}
Unlike blind participants, we found that low vision participants developed strategies for cooking and kitchen adaptations to better utilize their residual vision. In this section, we report low vision participants' unique challenges and strategies in the kitchen. 

\begin{figure}
  \includegraphics[width=\textwidth]{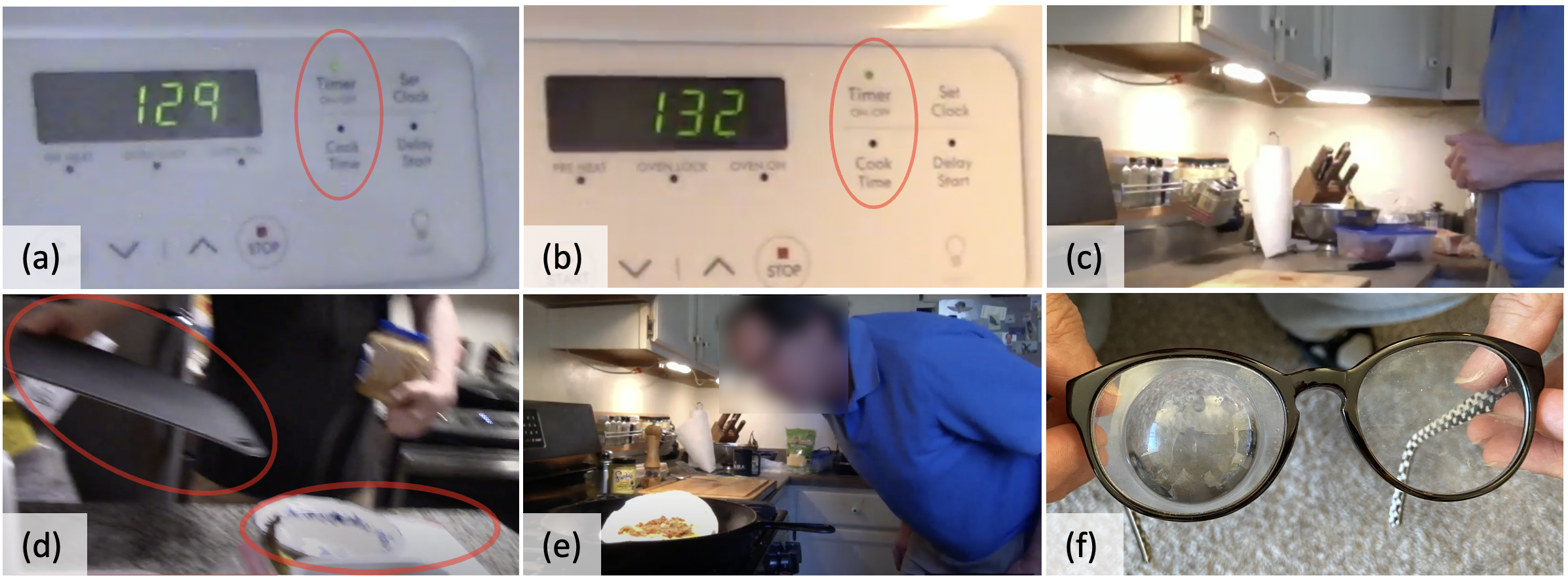}
  \caption{(a) The timer indicator light on oven control panel was more noticeable when the cooktop light was off, compared to (b) when the cooktop light was on; (c) Jacob installed a cabinet light over the countertop area to provide sufficient light; (d) Derek used black and white cutting boards for generating high contrast; (e) Jacob looked close at his food in the hot pan; (f) Judy's magnifying glasses.}
  \label{fig:low_vision_strategies}
    \vspace{-3ex}
\end{figure}

\paragraph{\textbf{Adjusting the Lighting.}}
Four out of six low vision participants relied on sufficient lighting to see things in the kitchen. Most participants turned on all the lights during cooking. Some even installed additional lights. For example, Jacob installed an under-cabinet light to light up his countertop (Figure \ref{fig:low_vision_strategies}c). With suitable lighting, \textbf{most low vision participants relied on the light reflection to identify kitchen elements and food ingredients}, including locating metal kitchen elements (e.g., faucet), distinguishing the edges of furniture, and gauging the amount of oil in the pan. However, the light reflection was not precise enough to support fine-grained kitchen activities. For example, Derek reported he placed a bowl right on the edge of the table by following the reflection, however, it then flipped over on the floor.

Moreover, \textbf{too much light can also cause low brightness contrast, which adversely affected low vision people's cooking experience} (Judy, Jacob). For example, Judy found it more difficult to see the status of the stove indicator lights when the cooktop light was on than having it off (Figure \ref{fig:low_vision_strategies}a and b). 
Since the intensity of the cooktop light was not adjustable, the cooking environment ended up being either too bright or too dark for low vision participants. As Jacob commented, \textit{``When I look at this [stainless steel pan], this light bouncing off. It's very, very bright, that pan is very, very bright to me. If I turn [the cooktop light] off, the pan is too dark, I can't even see anything going on, so I got to compromise and deal with that light.''}


\paragraph{\textbf{Creating Color Contrast.}}
Most low vision participants stressed the importance of color contrast in the kitchen. Participants encountered many cooking challenges due to the low contrast in the kitchen, such as reading text labels, cutting, checking food doneness. 
For example, when cutting, Judy struggled with seeing food with low contrast to the cutting board (e.g., white onions on white cutting boards), \textit{``All we have is white cutting board... I see the whiteness of the onions, but I couldn't tell the edges, or how many are right there.''} 

Telling the doneness of food can also be difficult if the food had low contrast to the pan (e.g., Jacob) or the contrast between cooked and raw food was low (e.g., Judy). 
To cope with this, some participants \textbf{observed the motion and shape changes of the food} during cooking. For example, Jacob checked the doneness of a fried egg by shaking the pan and observing the wiggling of the egg, and he checked whether the bacon was done by its curling shape.

Participants strived to achieve high contrast during cooking. Most participants
leveraged high contrast kitchen tools directly, such as measuring spoons with different colors (Chloe) or measuring cups with high contrast prints (Judy). \textbf{Some participants adopted unique strategies to create high contrast by themselves}. 
For example, Derek prepared two cutting boards---a black and a white one---and picked the cutting board with a good contrast to the food he cooked (Figure \ref{fig:low_vision_strategies}d). Interestingly, he used the cutting boards not only for cutting but also as a contrasting background for all kinds of cooking tasks, for example, putting a white bowl on the black cutting board to locate the bowl easily. 
Moreover, Jacob intentionally adjusted the order of the steps when making coffee to guarantee high contrast: instead of pouring coffee first to his black cup, he poured white milk first and then the coffee, so that he could easily see the liquid level due to the contrast with the cup. As Jacob described, 
\textit{``I have to look very closely [at the black cup]... If I pour this black coffee into here, it would be very difficult to see when is full. So I poured my milk in first, and I can see it's brown [after I pour the coffee], so I get some contrast with the brown coffee and the black cup. If it were black [without the milk] I would have to be really careful.''}

\paragraph{\textbf{Seeing Small Kitchen Elements.}}
Many low vision participants experienced low visual acuity, making it difficult to distinguish details.
Unfortunately, many kitchen elements and tasks involved seeing small details, such as reading ingredient descriptions, distinguishing food status. 
We observed that without using assistive tools, most low vision participants had to move very close to see various kitchen elements, including hot food. Jacob complained about bending over and putting his face close to the pan too often to see how well the food was cooked (Figure \ref{fig:low_vision_strategies}e). 
Locating and tracking small items (e.g., bottle lids) in the kitchen was also challenging for low vision participants. To address this, Derek usually placed small items next to big ones to generate size contrast.

\subsubsection{\textbf{Technology in the Kitchen.} }
\label{study1:tech_kitchen}
Participants used various technologies to assist cooking. Besides the various low-tech kitchen tools mentioned in Section \ref{study1:challenges_strategies_blv} (e.g., liquid level indicator), six participants mentioned using more advanced technologies, including AI technologies (e.g., Microsoft Seeing AI \cite{SeeingAI}) that recognizes text and kitchen objects, crowd-sourcing technologies to receive remote human guidance (e.g., Be My Eyes \cite{bemyeyes}), and smart speakers for basic appliance control (e.g., set a timer, read recipes, monitor temperature). Interestingly, Derek installed six smart speakers in his apartment, so that he could send commands and get responses from the speakers instantly whenever and wherever he needed to, especially in urgent situations. As he explained, \textit{``I want [the smart speakers] in close proximity, [so I can] just turn and talk to [them]. And I also have them in many locations for safety, in case I fall. That way, I can just yell from the floor. I can always make an outgoing call even if I'm stuck on the floor.''} 

Despite the usefulness, participants (e.g., Kim, Mary) stressed that \textbf{the slow (or non-real-time) feedback from current technology could not fulfill their immediate needs in the dynamic and busy cooking process.} For example, Kim mentioned that text recognition took too long when reading spices and teas, preventing her from completing tasks quickly. Similarly, Rene complained that she operated her microwave faster than Alexa did. As a result, many participants chose to complete the cooking tasks manually via touching and feeling (Mary and Rene). For example, Mary used her fingers to determine the bread yeast's temperature since her talking thermometer was too slow. 
Moreover, the current technology was not intelligent enough to adapt to the users' cooking ability. 
For example, Derek indicated that Alexa read recipes too fast to follow with his current cooking skills. 

Different from blind participants, low vision participants mostly using magnifiers in the kitchen. Three participants (Judy, Chloe, and Jacob) reported using magnifying glasses or handheld magnifiers to read food packages, kitchen appliance interfaces, and recipes during cooking. 
However, the traditional magnifier was not always sufficient for cooking activities because it cannot magnify things that were far away. For example, Judy mentioned not being able to use her magnifier to see the stove knobs since the stovetop blocked her, keeping her far away from the control panel. To address this issue, some participants (e.g., Chloe) used digital magnifiers, which captured the environment with a camera and magnified the digital video. Similarly, some participants (Jacob, Judy) used the smartphone camera to take a picture and enlarge the picture directly. However, capturing a clear photo of the object of interest was difficult for our participants. Derek thus suggested the camera providing adequate feedback to guide the user to take a clear photo.

Participants discussed their needs and expectations for future technology in the kitchen. To create more accessible kitchen appliances, six participants suggested embedding audio interfaces, including both voice control and audio feedback. For example, David suggested that the oven should 
notify the user of a certain temperature; Jacob wanted the stove and oven to periodically remind him when they were on. Most low vision participants also mentioned the need for big buttons with big and high contrast print on the appliance interfaces.
Additionally, Chloe and Judy wanted hands-free technology especially when both of their hands were occupied by a cooking task. For example, Chloe mentioned that she could not use her handheld magnifier when cutting. Interestingly, Hilary and Derek fantasized about an automated kitchen where a robot or some smart kitchen facilities could complete the cooking tasks for them automatically.
As Derek described, \textit{``If I could just put a food item [into] some little machine. Close [the] lid and tell it what to do, [and it would do it], would that be perfect.''}

\subsubsection{\textbf{External Help.}}
Besides the help from family, friends, and neighbors, participants also sought help from external services to facilitate cooking.
Derek and Judy used food delivery services that shipped cooked or pre-processed food to accelerate their food preparation, such as Meals on Wheels \cite{mealsonwheels} and Homechef \cite{homechef}. However, Derek pointed out that \textbf{the delivered food was not customized for customers' health conditions}, \textit{``Since I'm diabetic, it's a disadvantage because they include things that are not really the best for me.''} When he received the food, he usually tasted the food first to decide whether to eat it to control his carbohydrate intake. 


Derek also mentioned using home-care service to prepare a meal and organize the kitchen. However, \textbf{home-care workers had high turnover and were lack of training for working with PVI}, which largely diminished the service quality. For example, Derek complained about home-care workers changing the layout of his kitchen, making it difficult for him to find things,   
\textit{``It takes a while for [the home-care people] to understand that they have to keep things in the same place, same order...because you can literally move something from here over to here, and that can throw me off. And you know it takes me a long time to recover from, even though some of those simple changes in position.''}
Besides, \textbf{home-care service was vulnerable to the pandemic}. Derek had to stop the service with the increased severity of COVID-19. 



\section{Study 2: Interviews with Rehabilitation Professionals}

Rehabilitation training was a major way for PVI to learn strategies and technologies. Our first study with PVI showed that seven out of 10 participants had participated in such training. It is thus important to investigate this process to further understand how the training can potentially affect and shape PVI's experiences and skills in cooking activities. We conducted semi-structured interviews with six rehabilitation professionals with different certificates, such as occupational therapists (OTs) and teachers of students with visual impairments (TVIs). We explored their training strategies, recommended technologies for PVI, and their decision-making process, which helped us distill design insights for future assistive technologies in the kitchen from the experts' perspective. 
\subsection{Method}
\subsubsection{Participants}
We recruited six participants (4 females, 2 males), whose ages ranged from 30 to 54 ($mean=44.83, SD=7.47$). One participant had low vision, and five participants were sighted. All six participants had trained PVI in kitchen-related tasks, with years of experience ranging from 7 to 25 years. 

To recruit participants, we reached out to different blind and low vision organizations, including Vision Forward Association \cite{VisionForward}, Association for Education and Rehabilitation of the Blind and Visually Impaired (AERBVI) \cite{AERBVI}, and the National Federation of the Blind (NFB) \cite{NFB}, as well as the local hospital and clinics, including UnityPoint Health--Meriter Hospital and UW-Health low vision services \cite{Meriter}. They distributed our recruitment information and helped connect us with professional trainers. We also posted our recruitment information on the Reddit \textit{/rblind} channel. In addition, we used the snowball sampling strategy by asking our participants to share our recruitment information with other professional trainers.

We designed a survey on Qualtrics \cite{Qualtricssurveystudy2} to determine the eligibility of participants, where we asked about participants' training certificates, their training focus for visually impaired people, and their years of experience. Participants who had proper training certificates and had experience training PVI in kitchen tasks were considered eligible for our study.  

Our participants presented a good coverage of different types of trainers and organizations: three participants were occupational therapists (OTs) from non-profit organizations (Rachel and Lyla) or local clinics (Megan); two participants (Maria and Michael) were teachers of students with visual impairments (TVIs) from regular schools; and one (Kevin) was a chef coach for PVI from a cooking school. Table \ref{tab:OT_demo} detailed participants' demographic information and their training experiences.

\subsubsection{Procedure}
Our study consisted of a  single session that lasted one to two hours. All interviews were conducted remotely via Zoom. Our interviews included three phases: demographic information and general training experiences, kitchen-related training, and vision for technology. Besides interviews, we also observed three rehabilitation training sessions to thoroughly understand the training process. We described the details of each phase below. 

\begin{table}[t]
\small
\centering
\caption{Demographic information of the six rehabilitation professionals in Study 2.}
\begin{tabular}{C{1.5cm}C{1cm}C{1.5cm}C{1.5cm}C{4.5cm}C{1.5cm}}
\toprule
\textbf{Pseudonym} & \textbf{Age/{\newline}Gender} &\textbf{Visual {\newline}Ability} & \textbf{Title} & \textbf{Affiliation} & \textbf{Years of {\newline}Experience}\\ 
\midrule
Rachel & 45/F & Sighted & OT & Vision Forward Association & 8\\ 
\midrule
Lyla & 48/F & Sighted & OT & Vision Forward Association & 13 \\
\midrule
Megan & 54/F & Sighted & OT & UnityPoint Health--Meriter Hospital & 25\\ 
\midrule
Maria & 43/F & Sighted & TVI & Trevor-Wilmot Consolidated Grade School District & 11\\ 
\midrule
Michael & 49/M & Low Vision & TVI & Milwaukee Public Schools & 11 \\
\midrule
Kevin & 30/M & Sighted & Chef Coach & Highill Foods; UAC Foods & 7\\ 
\bottomrule
\end{tabular}
\label{tab:OT_demo}
\vspace{-3ex}
\end{table}

\textit{\textbf{Demographics and General Training Experience.}} In this phase, we obtained participants' demographic information, including age, gender, visual condition, educational level, and years of experience. Moreover, we asked about participants' general training experience, including their target training population (e.g., specific visual conditions, age range), types of tasks they focused on (e.g., navigation, cooking, technology use), and the format of training they provided (e.g., one-on-one sessions, group sessions). 


\textit{\textbf{Kitchen-related Training.}} We then focused on participants' training experiences with kitchen-related tasks. We asked participants about their general training procedure (e.g., training format, length of the training, and training content) and how they came up with training plans for each client. We then asked about the common cooking tasks that they focused on in the training. 
For each cooking task they mentioned, we followed up on their training principles, cooking strategies they taught, tools or assistive technologies they recommended, and the rationales for their training decisions. Moreover, we asked about how participants kept track of and assessed their clients’ progress as well as how they followed up with their clients after the training.

\textit{\textbf{Vision for Technology.}} The last phase of the interview focused on technologies, where participants talked about the benefits and weaknesses of the technologies they recommended in the training as well as how they discovered these technologies. We also asked about participants’ opinions on more advanced technologies, such as AI-based technology, smart glasses, and robotics. By the end, we prompted participants to brainstorm the ideal technology they desired to assist PVI to complete kitchen tasks independently and safely.

\textit{\textbf{Observation of Training Sessions.}} In addition to the interviews, we observed three rehabilitation training sessions under the agreement of both the professionals and their clients. Two sessions were led by Rachel where one was centered on baking brownies and the other one was centered on cutting different vegetables. Another training session we observed was led by Megan which focused on making a tuna salad. The observation was conducted remotely via Zoom, where we asked the professionals to turn the video camera on to stream the training process. We also scheduled a follow-up interview with the rehabilitation professionals to ask specific questions based on our observation that were not covered in the general interview, such as why they conducted specific behaviors (e.g., putting their hands on the clients' hands) or why they recommended specific ways of using an assistive tool (e.g., making puffy paint into specific shapes to label kitchen elements).   


\subsection{Analysis}
We took notes during the study and video recorded the whole interviews and observation sessions. We transcribed the video via transcription services and analyzed the data using qualitative analysis \cite{saldana2021coding, braun2006using}. We first developed codes using open coding. Two researchers independently coded three identical transcripts and watched corresponding video recordings to code participants' behaviors and answers. After the two researchers compared and discussed the codes, a codebook was generated based on their agreement. One researcher then coded the remaining transcripts and videos based on the agreed codebook. If new codes emerged, the codebook was iterated and updated upon researchers' agreement. We then derived themes using axial coding and affinity diagram. Since occupational therapy is well-structured process, the themes were generated using a deductive approach \cite{braun2006using} based on the standard training stages identified in prior work \cite{smallfield2020occupational, kaldenberg2020occupational}. When initial themes were identified, researchers cross-referenced the original data, the codebook, and the themes, to make final adjustments. Our analysis resulted in seven themes (Table \ref{tab:OT_themes}). 

\subsection{Findings}

\subsubsection{\textbf{Training Organizations and Their Target Clients}.}
Our participants were from different types of organizations and focused on different visually impaired populations. Among the three participants who were OT, Megan was the only one from a clinic setting, UnityPoint Health--Meriter Hospital. She mostly worked with clients who experienced acquired visual impairments caused by non-visual diseases or disorders, such as stroke, brain injury, multiple sclerosis, and concussion. 
While mostly working with older adults (50+ years old), she sometimes trained adults who were in their 30s or under 30. Her therapy primarily focused on activities of daily living (ADL) and specifically, on reading, writing, navigation, and cooking.

The other two OTs (Rachel and Lyla) were from a non-profit organization, Vision Forward Association. They worked with people from all age groups with congenital and acquired vision loss, and most clients were older adults (65+ years old). They had trained participants with various visual conditions, including macular degeneration, glaucoma, cataracts, diabetic retinopathy, retinitis pigmentosa, and Stargardt disease. Some clients also experienced other health issues, such as memory issues, mental health issues, and mobility issues (e.g., arthritis, Parkinson's disease).

Our participants also included two TVIs (Maria and Michael) from regular schools. Michael worked with visually impaired students from K-4 to high school. His training focused on daily living skills by following the individualized education program (IEP) for each student. 
Maria worked specifically with high school students (14 to 18 years old) who were legally blind. Besides following the IEP, she also provided expanded core curriculum (ECC) instruction \cite{sapp2010expanded} to students with visual impairments, which was a type of instruction specialized for students with visual impairments, incorporating nine core skills into students' academic lessons, such as independent living skills, orientation and mobility skills, and social interaction skills. 

Lastly, Kevin was a chef coach from a professional cooking school. His clients included adults (20+ years old) who were blind or had low vision. His training primarily centered on kitchen-related activities.

\subsubsection{\textbf{Training Plans.}} Due to the different working environments and target clients, participants adopted different training plans. We found that participants offered various training formats and implemented different levels of adjustments to the training plan according to their clients. We reported how participants planned and adjusted their training procedures below. 

\begin{figure}
  \includegraphics[width=0.9\textwidth]{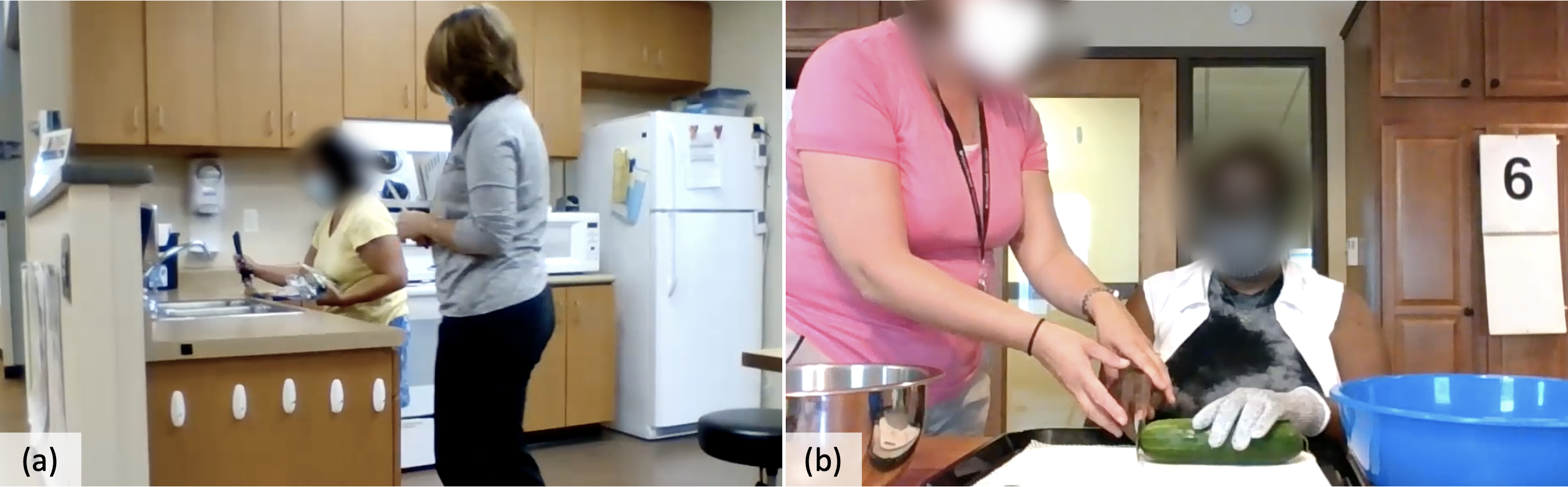}
  \caption{(a) One-one-one training in a standard kitchen environment; (2) Hand-over-hand instruction in a cutting task.}
  \label{fig:instruction}
  \vspace{-3ex}
\end{figure}

\paragraph{\textbf{Training Format.}}
Participants reported three training formats: \textit{one-on-one training}, \textit{small group training}, and \textit{large group training}. All participants but Kevin provided one-on-one training, where they met one client at a time, individually guiding each client through different training tasks (Figure \ref{fig:instruction}a). This training format enabled the professionals to better understand the needs of each client and tailor the training sessions accordingly. 

One-on-one training could take place either in a standard training space hosted by the training service or at the client's home. Rachel and Lyla supported at-home training. This type of training allowed them to observe clients' living environment, such as their kitchen setup and tools or technology they already had, and come up with a more suitable training plan. For example, Lyla explained that home visits helped her build a more accurate picture of the clients' needs as the clients' self-report might not reflect their actual situation: \textit{``I like doing home visits because, a lot of times, [my clients] will say that they're doing fine in an area that they're really not. When I can get in and actually observe them, I see that they are struggling with appliances [that they did not report having problems with], for example.''}


While only Megan provided one-on-one sessions exclusively, four participants (Rachel, Lyla, Maria, and Michael) also offered small group training with fewer than five clients per group. Small group sessions took place when the professionals observed some common interests or challenges from multiple clients. 
Although the small group sessions could boost training efficiency, they only applied to clients who were good learners and good team members.
As Rachel described, \textit{``if [clients] progress quite a bit through their individual training and want to continue to learn some techniques, and I have a sense that they're good team members, I might consider them for group cooking experience.''} However, the small group sessions were hindered by the COVID-19 pandemic: \textit{``We used to have group cooking sessions, where we focus on, for example, using small appliances. Because of COVID, our group cooking sessions are kind of on hold'' (Rachel).} 

Unlike other participants, Kevin adopted a unique large group training format at the professional cooking school. A large group was composed of over 50 clients with similar visual abilities (i.e., blind or low vision). Interestingly, to better track each client's progress, Kevin usually broke the large group down into groups of five, which ensured that the clients in the same group could progress at a similar pace. With this format, Kevin was able to implement a standard procedure to ensure that all clients achieve equal learning outcomes efficiently.

\paragraph{\textbf{Training Procedure and Timelines.}}
A training program usually consisted of multiple sessions that focused on different cooking activities (e.g., kitchen safety, use of kitchen appliances). The professionals thus needed to plan the training for each client, such as the number of training sessions and the focus of each session. We found that participants followed different approaches to design training plans and provided different levels of customization. 

Most participants customized their training plans for the clients. For example, Rachel and Lyla from Vision Forward usually planned multiple-session training for kitchen activities 
with each session lasting one to two hours. They customized their training plans based on the clients' abilities, goals, and performance. Before the training, an assessment was conducted to build a basic understanding of the client’s background and identify their goals. A customized training plan was then generated, including how many sessions were needed, what tasks to focus on, and what types of technologies to recommend. As the training proceeded, the OTs also adjusted the training plan dynamically according to the clients' progress and preferences. For example, if they noticed their clients had difficulty learning a specific task, they would repeat the same training content in the next session, sometimes using new training strategies. Rachel gave an example: \textit{``If they're focusing on stovetop use, for example, and they're having some challenges with it, we would do another session the next time before we move on to another skill area … and I can break the training down into a subset of [small] tasks [to help them learn better].''}


The TVIs customized their training plans by following the IEP, where they set up learning objectives with each student at the beginning of a year, along with specific evaluation criteria to reflect their training outcomes. As Michael described: \textit{``If [the students] need help in a certain skill in daily living, then that's a goal that we would write for them in their IEP and create objectives. That goal could be being able to operate a microwave or an oven independently, say, two out of three trials with an 80 percent success rate.''} At the end of the year, the TVIs also conducted an annual discussion with the students to review their progress, goals, and potential adjustments to the goals for the next year. While offering high flexibility for training content, the TVIs had to follow the general academic schedule of the school. Specifically, a student received a training session once a week or once every few weeks outside of school hours with each training session lasting 30 minutes; and the training had to be terminated when the students graduated.  

Megan from the clinic setting provided a lower level of customization. She usually followed a general routine, covering a few common goals and skills during the training, such as reading package information, measuring, and cutting. However, she was still open to adjusting her plan if the client showed interest in a particular goal.  


Unlike other professionals, Kevin from the cooking school followed a standard training plan with a strict timeline since he had a large group of students. No customization was provided during his training process. The training lasted 13 to 14 weeks, with every two weeks focusing on a specific cooking task, such as distinguishing utensils and identifying spices. The clients needed to pass a test every two weeks to move on to the next cooking topic. 

\subsubsection{\textbf{Assessment Methods.}} During the training, participants used different methods to assess their clients' progress and needs to adjust the training plans. We reported the assessment methods applied at different training stages.

\paragraph{\textbf{Pre-Training Assessment.}}
A pre-training assessment was usually conducted to evaluate the clients' ability and needs to decide the initial training plan. One of the most common pre-training assessment methods was interviewing the clients, which was used by five participants (except Kevin). During the interview, the trainers asked about the client’s visual and other (dis)abilities, prior experiences in the kitchen, and their training goals. Our participants also adjusted their assessment methods according to the client's abilities. For example, Rachel mentioned simplifying the interview questions and involving family members if her client had cognitive disabilities: \textit{``For somebody who is cognitively impaired, I am probably going to be very basic and simple with my questioning, and I may even be asking a family member or somebody the questions instead of the client.''}

Besides the clients' basic information and needs, our participants also assessed the client's family involvement and expectation. As Michael emphasized, \textit{``[The training plan] depends on the dynamics of the student and what their expectations are at home. Some parents look at their child being so disabled and [think that] they can't do anything on their own, [whereas] other parents may really encourage them to try to be independent.''}

Three participants (Michael, Maria, and Kevin) used the method of observation during the pre-training assessment. Specifically, Michael and Kevin designed small tests to gauge a client's familiarity with certain kitchen tools. For example, Michael gave his clients a tactile diagram of a stove and observed how they used the controls on it (Figure \ref{fig:tactile diagrams}). 
To better observe the clients, Maria conducted a trial session with the client before the formal training. In the trial session, she provided some basic training for a specific task and observed the client's performance to evaluate their current cooking skills and ability to learn. Maria described one of her trial lessons that focused on measuring: \textit{``I'll do like just a mini-lesson, like can you measure something with a measuring spoon accurately? And I might just use a bowl of rice and a measuring spoon and see if they know how to level off the spoon. So, if I'm doing those mini-lessons, I'm gonna see how quickly they are picking up.''}

\paragraph{\textbf{During-Training Assessment.}}
Participants reported different methods to assess clients’ progress during the training. All participants mentioned measuring clients’ performance (e.g., the speed of completing a kitchen task, the success rate) and their psychological status (e.g., comfort level, confidence). For example, Rachel evaluated clients' comfort level by observing whether they were hesitant and whether they asked many questions: \textit{``I just watch how they perform the task. Most clients will tell me if they're nervous about something or they will hold back and be very careful about how they're doing, so if they're asking fewer questions and performing it well, that tells me a lot about their comfort level.''}

Furthermore, Lyla and Megan used progress notes every three months (or 10 visits) to document their clients’ progress. For example, Megan used a Patient-Specific Functional Scale (PSFS) for the client to rate their current abilities on various activities or tasks, allowing her to adjust the training plan accordingly. 


Different from all other participants, Kevin, the chef coach, evaluated his clients' performance via biweekly tests. He then graded the clients' cooking skills from different aspects. Only clients who passed the test could continue their training for new cooking topics.
\begin{displayquote}
 \textit{``At the end of two weeks, we have tests, and we make sure our students are over 75 percent. It's like grading points, safety should be 20 percent, identification should be 10 percent, success in your mini test should be 10 percent, and ability to work in a group just like teamwork is 10 percent.''}
\end{displayquote}

\paragraph{\textbf{Post-Training Assessment.}}
Not all participants conduct post-training assessments. Only Rachel and Lyla followed up with their clients by phone interviews to ask about their current situation and check whether they need more training. The interview took place one or two months after the training.

Interestingly, Kevin conducted a final test after the entire training process. The final test covered different cooking tasks that the clients had learned during training, \textit{``We actually ask them to cook a full meal. They get to cook pastries, soup, baking, so just different kinds of categories. You have to make these four meals, and there are some that are given in a group and some must be done individually.''} If his clients were unable to pass the final test, they had another four-week window where they could receive more training in order to pass the final test. 


\subsubsection{\textbf{Training Principles.}} To guarantee clients' learning outcomes, participants followed some common principles throughout the training process. We reported the principles below.

\paragraph{\textbf{Ensure Safety.}} \label{ensure_safety}
All participants emphasized the importance of safety in the kitchen and used different approaches to protect clients' safety during the training. For example, Maria and Kevin usually discussed the potential safety risks of cooking with their clients at the beginning to mentally prepare them for the underlying hazards. Moreover, Michael asked his clients to first navigate around and familiarize themselves with the kitchen before the formal training, 
thus protecting them from potential navigation risks (e.g., bumping into table corners). As Michael said, \textit{``For our students, it's really [keeping] them under close supervision, [exploring] on their own, letting them go around the kitchen, and seeing or feeling to explore the different components of the kitchen.''} 

When training the PVI in relatively dangerous cooking tasks (e.g., cutting, using hot appliances), instead of rushing into the task, some participants (e.g., Maria, Rachel) encouraged their clients to first freely explore the specific tools or appliances. For example, in the training for oven use, Maria had her clients touch around the oven and feel the inside before turning it on to better understand the appliance structure and interface. Another example was during the observation of Rachel's session on cutting, she first presented a range of cutting tools. While the client felt and identified each tool, she would explain the function of each one before asking the client to start cutting.

Cooking strategies and tools were also introduced by our participants to facilitate PVI's safety in the kitchen. For example, protection tools (e.g., cut-resistant gloves) were always recommended, and hot appliances should be turned off when moving food in and out of the appliances. Moreover, when organizing kitchen elements, Michael suggested his clients put heavy items on a lower shelf instead of a higher one to prevent the risk of heavy items falling from high places:\textit{``Putting cans on a lower shelf, so if they fall, they won't fall on your head. Putting boxes or things that are lighter in weight and putting those more at the top just for safety concerns.''} 

\paragraph{\textbf{Maximize Vision Use for Low Vision Clients.}}
When training low vision clients, all professionals encouraged their clients to leverage their residual vision.
As Megan mentioned,\textit{``If [the clients] are low vision and they do have some residual vision, then we try to do as much as we can visually. Depending upon the range of their vision, the more vision they have, obviously the more visual support they'll get.''} To maximize vision use, participants recommended kitchen tools with bright colors or strong color contrast. For example, Rachel encouraged her clients to prepare a white and a black cutting board so that they can choose the one with high contrast according to the food they need to cut.

\paragraph{\textbf{Encourage the Use of Other Sensory Channels.}} 
For clients who are blind or have progressive low vision, our participants encouraged the use of other sensory channels in the kitchen, such as touching, smelling, and hearing. 

Participants recommended different ways of touching. Barehand touching was recommended in many kitchen tasks, such as cleaning dishes, measuring, and using appliances. For example, when teaching measuring, Maria instructed the clients to put their fingers at the top of the measuring tool to feel the level of the ingredient. However, when dealing with hot food, all participants recommended using tools as "an extension of hand" for safety (Rachel). For example, Rachel suggested using a long handle wooden spoon to tap around and locate food in a hot pan without getting burned. 

Some professionals encouraged their clients to smell or listen. For example, Kevin taught his clients standard ways to smell and identify ingredients. Meanwhile, Michael asked his clients to listen to the change of the sound pitch as the liquid level increased when pouring. 



\paragraph{\textbf{Practice at Home.}}
Besides learning cooking techniques during the training, all participants stressed the importance of practicing cooking skills at home after training. However, participants expressed divergent opinions on when to start practicing. Three participants (Rachel, Lyla, and Maria) encouraged the clients to practice at home at any time so that they could explore their preferred way of cooking instead of following strict rules or instructions. As Maria described: \textit{``I am very open to having [my clients] try at home. I am a person who [wants] you to do something for the joy of it. And if I make it very regimented, I'm going to rob the joy of it almost immediately. I don't find that to be successful as a strategy for learning almost anything''}. As a TVI, she even talked to her students' parents, encouraging them to let their children help in the kitchen at home. Moreover, Lyla expressed the benefits of practicing for clients with other disabilities, for example, enhancing the memory of older adults with memory loss and increasing the muscle tolerance of people with physical disabilities. As she mentioned, \textit{``I tell [my clients] to go over all the things that we went over that same day, don't even wait till tomorrow and do it today to lock it in. For a lot of our clients, because they're older, they do have short-term memory loss, so they have to practice.''}

However, Megan and Kevin were against practicing at home until the client could use all the techniques safely and independently. Without supervision, a visually impaired client at the early training stage could face safety risks in the kitchen when practicing at home. As Megan said, \textit{``Anything with the stove, heat, knives, I want to make sure that [my clients] are safe and independent using it in a kitchen before I have them practice at home. But once they are safe and independent, then yes, they can practice at home.''}


\paragraph{\textbf{Involving Specialists from Other Fields.}}
To provide more sufficient and professional training, our participants also involved other specialists in specific kitchen tasks. For example, Michael asked an O\&M instructor to be present to help the clients navigate the kitchen: 
\textit{``Another person, like an orientation and mobility instructor, may be involved and they'll help [the clients understand] the ergonomics of the kitchen or the room they'll be working in, so they know where the sink is and where that towel maybe.''}
Furthermore, participants also referred their clients to other specialists or organizations when their service could not fully address the clients' needs. For example, Rachel made referrals when she noticed that her client had cognitive or mental issues. 

\subsubsection{\textbf{Training Instructions and Strategies.}}
Participants used various strategies to make the training more effective. All participants provided verbal instructions to explain different cooking skills. When the clients practiced the skills, participants also observed their behaviors and provided verbal comments and suggestions. 

Some professionals (e.g., Rachel, Megan, Michael) adopted hands-on instructions, using their hands to guide the clients directly. Rachel described two types of hands-on instructions: \textit{hand-under-hand} and \textit{hand-over-hand} guidance. When teaching a new cooking task, she used hand-under-hand guidance, where she put her hand under the client’s hand to conduct the task. The client could thus feel her hand postures and motions in the task. For tasks the client was already familiar with, she switched to hand-over-hand guidance, putting her hand over the client's hand and only providing guidance when necessary (Figure \ref{fig:instruction}b). As Rachel described, 
\begin{displayquote}
 \textit{``For a lot of people with vision loss, if they're doing something for the first time, hand-under-hand is helpful because I’m doing the tasks myself, and they have their hands on me so they can get an understanding of it. With hand-over-hand, they have a little bit more control and they're doing the tasks themselves but I’m guiding them a little bit.''}
\end{displayquote}

\begin{figure}
  \includegraphics[width=\textwidth]{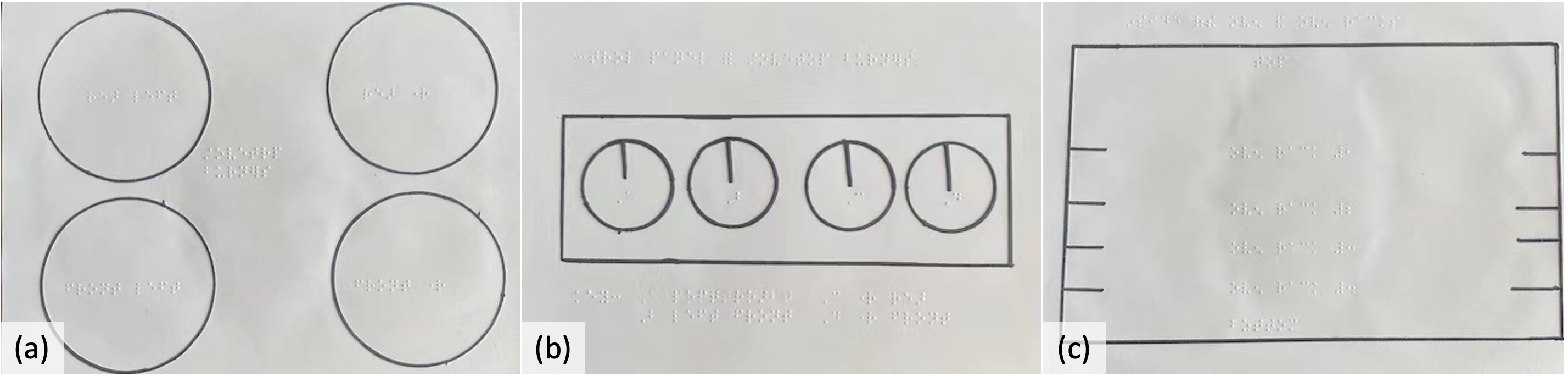}
  \caption{Tactile Diagrams with Braille: (a) the burner layout on a stove; (b) a dial interface on a stove; (c) a top-down cross section of an oven with a rack.}
  \label{fig:tactile diagrams}
  \vspace{-3ex}
\end{figure}

Notably, Michael leveraged tactile diagrams in his training, which enabled his clients to feel some visual components in the kitchen. For example, he created tactile diagrams with braille labels to demonstrate the stove and oven interfaces, including the burner layout on a stove (Figure \ref{fig:tactile diagrams}a), a dial interface to control the stove (Figure \ref{fig:tactile diagrams}b), and a top-down cross-section of an oven with a rack (Figure \ref{fig:tactile diagrams}c).
\begin{displayquote}
 \textit{``I did different diagrams---an overhead view of what the burners looked like on the stove and another of the inside of the oven so that the student understood there were racks in the oven. I did one more diagram of the top of the surface where the dials were located on the stove and then gave [my client] that to explore with his fingers and hands.''}
\end{displayquote}

\begin{figure}[b]
  \includegraphics[width=\textwidth]{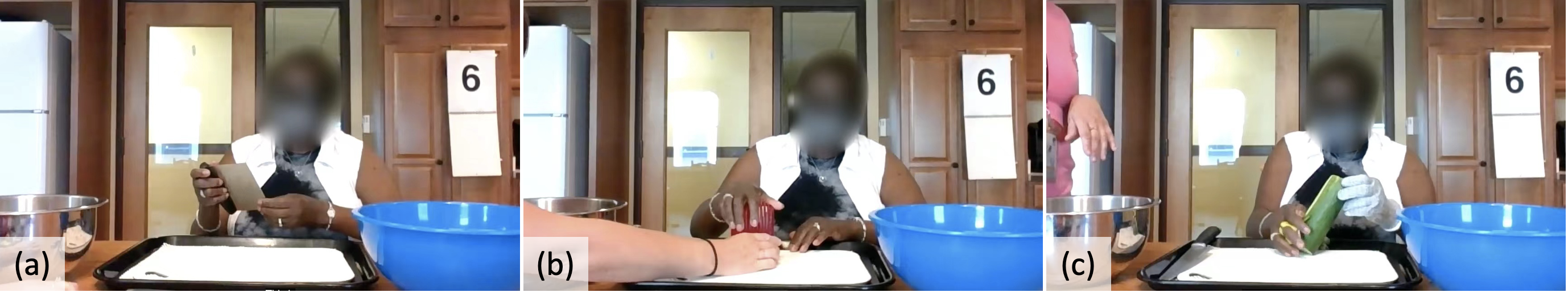}
  \caption{Trial-and-error with different cutting tools: (a) a scraper to cut cakes and brownies; (b) a plastic cutting mold to cut food in uniform pieces; (c) a palm peeler to wear on a finger to peel a cucumber.}
  \label{fig:trial-and-error}
\end{figure}

When recommending tools or technologies to clients, most participants (Rachel, Lyla, Megan, Maria) employed a trial-and-error method. They asked the clients to try a series of tools and picked the most suitable one based on their preferences. For example, in a cutting task, Rachel provided multiple cutting tools for her clients to try out, such as a scraper to cut cakes (Figure \ref{fig:trial-and-error}a), a plastic cutting mold to cut food in uniform pieces (Figure \ref{fig:trial-and-error}b), and a palm peeler to wear on the finger to peel vegetables (Figure \ref{fig:trial-and-error}c).

\subsubsection{\textbf{Training Content.}} \label{training content}
Participants identified seven common skills that were important in cooking and/or challenging to visually impaired clients. They included five cooking-specific skills---cutting, measuring, pouring/transferring, dealing with hot appliances, and checking food doneness, as well as two general skills---labeling and reading. 
We reported the strategies suggested by participants for each skill. 

\begin{figure}[b]
  \includegraphics[width=\textwidth]{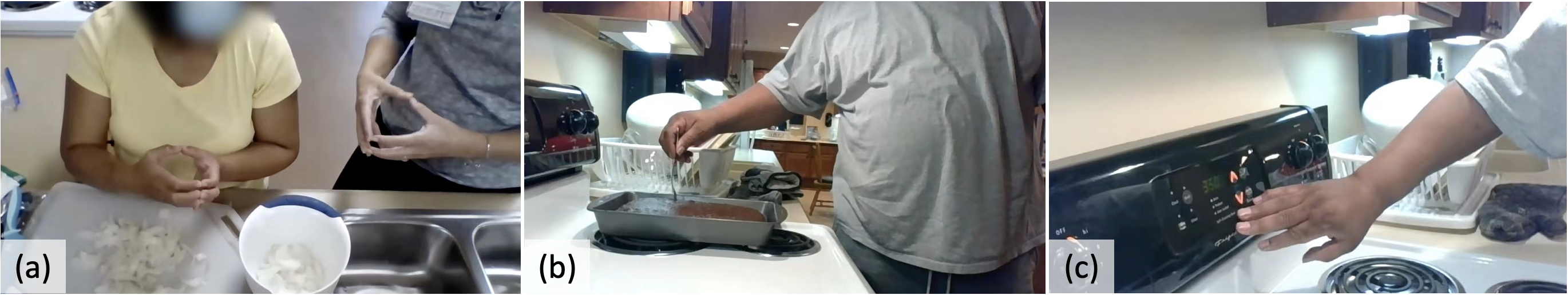}
  \caption{Strategies for basic cooking skills: (a) Use hands to form a circle to help measure the amount of food needed; (b) Use a fork to determine the food doneness; (c) Make arrow shapes with puffy paint to label the control buttons on an oven.}
  \label{fig:training content}
\end{figure}

\begin{itemize}
    \item Cutting: (1) When holding food, curl the fingers under so that the nails touch the food and the flattened knuckles are exposed to the blade to prevent cutting fingertips; (2) Use alternative tools that are easier to control and provide more feedback, for example, a palm peeler that allows bigger contact area between the client's hand and the vegetable (Figure \ref{fig:trial-and-error}c); (3) Wear cut-resistant gloves; (4) For low vision clients, use a cutting board with high color contrast to the food (e.g., cutting onions on a black cutting board). 
    \item Measuring: (1) Use hands to measure food amount, for example, forming a circle with two hands to collect food together and feeling the food amount based on the size of the circle (Figure \ref{fig:training content}a); (2) Use measuring spoons or cups with tactile marks (e.g., Braille, bump dots); (3) Use a talking scale to measure precisely.
    \item Pouring liquid/Transferring food: (1) Pour liquid over a sink in case of spilling; (2) Listen to the pitch of pouring sound for liquid -- the pitch of pouring sound rises as the liquid increases; (3) Touch the top of the container to feel the food or liquid reaching the top.
    \item Dealing with hot appliances: (1) Wear or install protection tools (e.g., oven mitts, oven rack guards); (2) Keep appliances turned off or unplugged when moving food in and out; (3) Label appliance interfaces with tactile labels properly (see specific strategies for labeling below). 
    \item Checking food doneness: (1) Use a talking thermometer to check food temperature; (2) Touch the food with a fork to feel the texture and tenderness (Figure \ref{fig:training content}b).
    \item Labeling: (1) Use semantic tactile labels, such as Braille labels and symbols/letters made of puffy paint (Figure \ref{fig:training content}c) to reduce users' learning curve and cognitive load. (2) Only label the most important or commonly used locations (e.g., the one-minute button on a microwave) and avoid over-labeling.
    \item Reading: (1) Use a barcode scanner to read food packages, such as the name of the food and nutrition information; (2) Leverage AI technology (e.g., Optical Character Recognition) to automatically recognize text in the kitchen (e.g., recipes, food packages) and receive audio feedback; (3) Adopt more accessible recipes, such as Braille recipes and auditory recipes.
\end{itemize}

While the above strategies focused on the kitchen context, participants emphasized that the strategies for labeling and reading could be generalized to other daily tasks, such as organizing closets, labeling medicines, and reading mails. 


\subsubsection{\textbf{Technology Recommendations.}} 

Participants recommended and provided training on various technologies to support kitchen activities, ranging from low-tech tools to advanced intelligent technologies. Low-tech aids included electronic talking tools (e.g., talking thermometer, talking scale), labeling tools (e.g., Wikki Stix, bump dots), magnifiers (e.g., wearable optical magnifier, handheld magnifier with LED lights), and protection tools (e.g. cutting gloves, oven rack guards). In terms of advanced technology, participants mentioned recognition technology (e.g. OrCam \cite{OrCam}, Microsoft Seeing AI \cite{SeeingAI}) to read food packages and recipes, vision enhancement smartglasses (e.g. eSight \cite{ESight}, IrisVision \cite{IrisVision}) for low vision people to see better in the kitchen, and smart speakers (e.g. Alexa \cite{Alexa}, Google Home \cite{GoogleHome}) to control kitchen appliances. 

However, we found that all participants tended to recommend low-tech tools rather than high-tech devices due to multiple reasons, including cost, learnability, stability, and maintenance. First, advanced devices often were more expensive and the cost could not be covered by insurance. As Rachel indicated, \textit{``A lot of our clients [are] coming through insurance, and the insurance won't often pay for adaptive tools, which is unfortunate.''}
Second, the maintenance of high-tech devices, such as charging and updating, frustrated some visually impaired clients since they did not know when and how to charge the device (Rachel). 
Moreover, learning advanced technology could also be difficult for PVI, especially for older adults and people who were not tech-savvy (Rachel, Lyla, Maria, Michael). The lack of stability of some high-tech devices further diminished PVI's experience. For example, Michael mentioned that the OCR technology failed to recognize certain fonts and images that were not clear enough. 

Due to the drawbacks of current technology, all participants believed that low-tech tools were more suitable to be recommended to a larger population. As Megan expressed, \textit{``Low tech devices are easy to operate, inexpensive, clinics have them, patients can get them. High tech tools like smart glasses are very expensive and they take some learning [curve] as well, and so out of reach for the majority of people that I have worked with.''}
Additionally, most professionals had limited resources to learn new technology, which also prevented them from recommending the latest technology to clients. Compared to other participants, Rachel and Lyla from Vision Forward were more familiar with advanced technologies (e.g., OCR, smart glasses) since Vision Forward had a specialized tech department to track technology updates and provide demonstrations. As Rachel described, \textit{``We have an assistive technology department and assistive technology experts here. They are up on the latest of everything related to vision loss, so if a client has a need or an interest in a certain tool, I can just show them how to use it.''}
In contrast, other participants had to do their own research to keep up with the latest technology, however, their time to do so was limited. As Michael indicated: \textit{``We have days off where we do professional development. Sometimes our vision department will get together and do training on technology. A lot of times, you're on your own, and it's up to you to keep up with it.''}

Although all agreed that high-tech devices were challenging for older clients, participants expressed divergent opinions on young clients' technology preferences. Both Michael and Maria taught students with visual impairments. On the one hand, Michael indicated that his students did not prefer technology. Instead, they would rather use their hands or low-tech cooking tools in the kitchen because they were conscious of appearing different from their peers by using additional devices. \textit{``When you start bringing in technology or they're using things that look different from other people, that makes them feel uncomfortable and different'' (Michael).} On the other hand, Maria reported that her students were not concerned about looking different from peers. Instead, they were good at and passionate about technology and preferred to incorporate technology in their daily life, such as reading accessible textbooks on a laptop: \textit{``I would say the majority of my students thrive when allowed to use technology. I actually had a student just today tell me 'I want to take my Chromebook home because that's how I [access] my math book' I think my students thrive because they're usually better at it.''} 


\subsubsection{\textbf{Desired Technology for Cooking.}}
Participants expressed their vision on future technologies that can better assist PVI to cook safely and independently. 
Most participants (e.g., Rachel, Megan, Maria) expressed the need for technology to protect PVI's safety in kitchen tasks, such as cutting, using hot appliances, and navigating crowded kitchen environments. For example, Rachel suggested heat-resistant labeling tools for burners to prevent burning risks: \textit{``A lot of electric stoves are going towards a flat, sleek, and nice-looking surface. But for somebody with vision loss, they cannot see and tell where the burners are. We can't put any type of our traditional labels on the flat top surface because they will burn. It would be cool if a company would design [new labels] that can easily be just put down on the stovetop.''} Rachel and Megan also suggested designing technologies that can guide PVI during cutting, enabling them to cut food safely and uniformly.  

Four participants (Rachel, Michael, Megan, and Lyla) mentioned providing multi-modal cues to support cooking tasks for PVI. For example, Michael pointed out that the flat control screen on many appliances was inaccessible and suggested adding tactile controls, such as buttons, knobs, and dials. \textit{``We have flat top stoves and refrigerators now with TV screens on it. I think appliances should have dials and buttons that you push and where you turn manually, instead of all of these digital buttons and displays.''} Megan suggested incorporating screen readers to kitchen appliances to provide audio feedback, especially for novice PVI: \textit{``The first time they get their machine with the buttons, they're supposed to push the button and it explains what it does, and to know what the different icons at the bottom do and how to access the best out of them.''}

Megan and Lyla emphasized the importance of hands-free technology, such as wearable devices or IoT devices embedded in the kitchen, so that PVI did not have to hold a magnifier or a phone during cooking, especially when they needed both hands to complete a cooking task. Lyla explained that \textit{``I like hands-free. When people are in the kitchen, they are using both hands, so they are busy. They are carrying things, they're cutting, they're putting things in whatever.''}

Moreover, most participants (e.g., Rachel, Megan, Michael) highlighted the need for potential visual augmentations in the kitchen to assist low vision people. They suggested incorporating bright colors, high contrast, and suitable lighting in kitchen tools and appliances. As Rachel explained, \textit{``We always emphasize the contrast with opposite colors. Sometimes you'll see a display that has a gray number against a black background [on some kitchen appliances], and it makes it hard to see. So for somebody with some vision, we want bright colors and opposite colors on displays.''} 


\subsubsection{\textbf{Training Challenges and Limitations.}}
Except for Kevin, who was confident that the cooking school addressed every aspect of PVI in the kitchen, all participants agreed that rehabilitation training could not address all difficulties that PVI faced in the kitchen and discussed the limitations and challenges in the training process. 

One general challenge was that the training was usually conducted in a relatively standard and simple environment so that skills learned in such an environment might not be easily transferred to clients’ own kitchen. As Maria described, \textit{``We don't have 65 bottles of spices. But, for a cook at home, that is a problem you're going to run into. Because I'm in a very controlled environment that is not like their house. I don't think I can truly mimic the real-life experience.''}

Clients' emotions could also pose barriers to effective training. Participants reported that their clients expressed fear and nervousness, and some even became emotional during the training. As Rachel mentioned, \textit{``[Vision loss] is hard to deal with if it causes [people] to change their role. If a person was a chef before, and now they lost their vision and they are like 'I can't cook, how can I use the stove?'''} Moreover, if a client experienced vision loss recently, they could be in the emotion of \textit{``grieving their vision loss''} and felt less ready and motivated to respond to training (Lyla). To help PVI deal with their emotion, both Michael and Rachel suggested seeking help from support groups or social workers. 

Lack of motivation could also hinder PVI's training progress. Compared to older clients, young clients were much less motivated to learn living skills since chores like cooking were usually taken care of by their parents. As Michael pointed out, \textit{``I think older people are more willing to ask 'how can I do this?' because it's something that they've always done [before becoming blind], and they want to find ways to continue the lifestyle that they've always known. I feel like for kids nowadays, so much has been done for them, so their need of wanting to learn how to do things isn't as much.''}

Another issue unique to students with visual impairments was their limited time for living skill training due to their heavy academic curriculum load. Since the IEP had to follow the general academic schedule, there's little time left for the students to participate in training for ADL:
\begin{displayquote}
 \textit{``Because with academics, there's so much [academic curriculum] you have to teach in a day. And you almost have to find another part of the school day to section off for this kind of [living skill] training. But then, what do you have to take away from their school day to fit that in? It's very challenging. It's [difficult to find] the time to do it when you have such a busy school day, and it's hard to fit that in'' (Michael).}
\end{displayquote}
\section{Discussion}

Our research contributed a thorough observation of PVI's cooking experience as well as the first investigation on the rehabilitation training process for kitchen activities. Unlike prior work that focused on completely blind participants \cite{kashyap2020behaviors, li2021non}, we conducted a contextual inquiry study with participants with different visual conditions---six low vision participants vs. four blind participants; four participants with acquired visual impairments vs. six participants with congenital visual impairments. Beyond confirming and triangulating the valuable findings about blind people from prior work \cite{kashyap2020behaviors, li2021non}, 
our study revealed more detailed cooking behaviors and uncovered different cooking challenges, strategies, and needs by people with different visual conditions.

Beyond observing how PVI cook in daily life, understanding how they learn cooking and the available rehabilitation resources was another important aspect of our research. We interviewed six rehabilitation professionals from different programs and observed some cooking training sessions. This study revealed the detailed training process (e.g., the training format, content, and principles) as well as the limitations of current training. While different services offered different training plans, we found that most professionals provided the flexibility to customize the training procedure based on PVI's abilities and needs. 

The two studies with PVI and rehabilitation professionals enabled us to develop a comprehensive understanding of PVI's cooking and training experience, thus triangulating their challenges and needs in the kitchen. Contextualized in prior insights \cite{li2021non, kashyap2020behaviors} and technologies \cite{mery2013automated, sato2014mimicook, ramil2021allergic, kim2022vision}, 
we discuss the gaps between rehabilitation training and PVI's cooking needs in daily life, as well as the technology implications to train and support PVI in the kitchen. 

\begin{figure}
  \includegraphics[width=\textwidth]{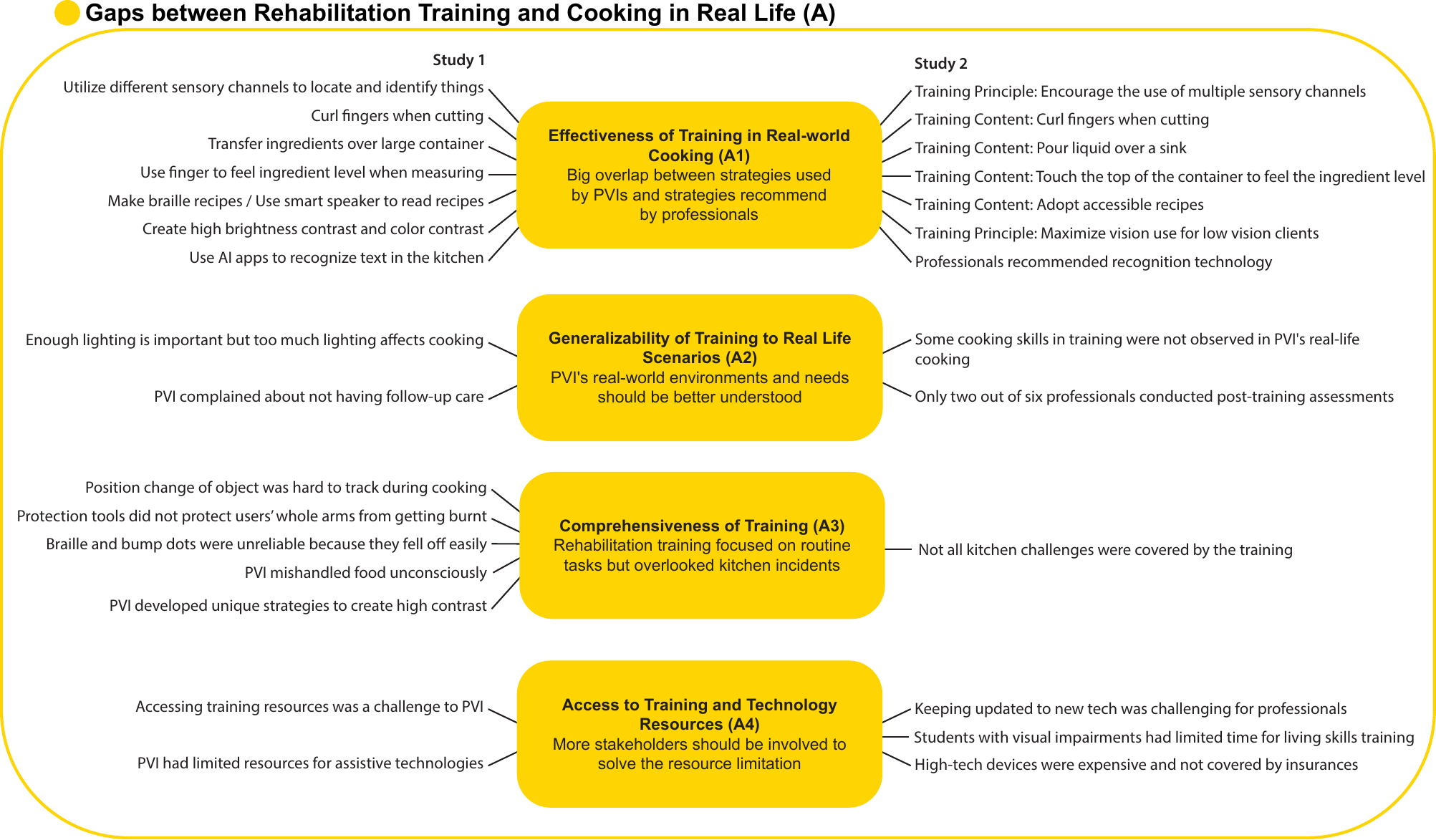}
  \caption{A diagram of the connections and gaps between rehabilitation training and PVI's real-life cooking experiences}
  \label{fig:connection_A}
    \vspace{-3ex}

\end{figure}

\subsection{Gaps between Rehabilitation Training and Cooking in Real Life}
\label{training-to-life}
Little research had investigated the rehabilitation training process and its impact on PVI via the lens of HCI. Our research explored the training process and strategies in different rehabilitation programs (Study 2) as well as how the training experience affected PVI's cooking in real life (Study 1). We discuss the connections and gaps between the training process and people's real-world cooking experiences; Figure \ref{fig:connection_A} shows a detailed diagram of the relationships.

\subsubsection{Effectiveness of Training in Real-world Cooking (A1).} Rehabilitation training is helpful and can largely affect PVI's cooking habits. In our study with PVI (Study 1), we observed that participants used a lot of cooking skills and tools they learned from the rehabilitation training. For example, Derek spoke highly of the George Forman Grill recommended by his occupational therapist and used it frequently. Moreover, the cooking tasks that the professionals focused on (Study 2, Section \ref{training content}) were all major challenges mentioned by our visually impaired participants in Study 1. We also found a big overlap between the cooking strategies and tools introduced by the professionals (Study 2) and the actual strategies adopted by PVI (Study 1) across all cooking tasks, for example, cutting with curling fingers and using white and black cutting board to generate high contrast. The consistent results between the two studies indicated the effectiveness and practicality of the training to be applied to real-world cooking. 

\subsubsection{Generalizability of Training to Real Life Scenarios (A2).}
While the training provides useful strategies and tools, some strategies recommended by the professionals (Study 2) may not be applicable to PVI's real-life cooking (Study 1). For example, using talking thermometer to check the food doneness was recommended by professionals, but no PVI adopted it in Study 1 due to the slow feedback (Mary). Thus, understanding the clients' actual cooking needs is important when preparing training content. Furthermore, since PVI's kitchen environments at home are usually more complex and challenging than the ideal training environments, the skills learned from the training may not be easily transferred to support PVI’s everyday cooking. For example, Jacob liked the professionals' advice to install enough lighting in the kitchen, but the glare on silver cookwares was detrimental to his cooking. Although post-training assessment was used by professionals to understand how their clients applied learned skills into their daily lives, only two out of six professionals in Study 2 conducted post-training assessments. PVI in Study 1 also mentioned no follow-up care was available when they struggled after training. As such, it is necessary for rehabilitation professionals to consider how to improve the generalizability of rehabilitation training to clients' real-world environments and needs, for example, increasing post-training, at-home assessments to understand PVI's real-life cooking experience.

\subsubsection{Comprehensiveness of Training (A3).}
The skills introduced in the rehabilitation training cannot cover all PVI's challenges in the kitchen. Six out of ten participants in Study 1 participated training, however, they still faced various difficulties or made mistakes during cooking. For example, some PVI had trouble tracking the position of utensils during cooking; some failed to detect the non-edible packet on pizza crust; and some mistook corn starch with baking soda. However, no professionals in Study 2 mentioned teaching their clients how to prevent and handle these errors quickly and safely in cooking tasks. 
Furthermore, we found some PVI discovered unique strategies that were not mentioned by professionals, such as pouring coffee into milk to create high contrast to avoid spilling. It would be helpful if the professionals can learn about these strategies developed by the PVI and further introduce them to other clients. The learning process between PVI and professionals can be bi-directional. 
Future researchers should explore how to facilitate the interaction and communication between the professionals and the PVI to refine the training content.


\subsubsection{Access to Training and Technology Resources (A4).} 
The limited resource access and time of both PVI and professional trainers prevent PVI from participating in rehabilitation programs. In Study 1, several PVI mentioned that they did not know how to get rehabilitation training and whether they were eligible for it (e.g., Judy). Even participants with training experience reported the difficulty of gaining sufficient training due to the limited resource in the training program (e.g., Dixie). 
In terms of technology, many advanced assistive devices were expensive and not covered by insurance, which may not be affordable to many PVI. This resource limitation was also reflected by the rehabilitation professionals in Study 2: professionals did not have enough time and resources to keep updated to new technologies,
so that their technology recommendations usually focused on low-tech tools. Moreover, for students with visual impairments, their training (i.e., IEP) had to make way for the general academic curriculum, which further limited their time for training on living skills. Unfortunately, the resource and time limitations could not be simply addressed by technology. More stakeholders in this field should be involved in the conversation, from the clients and professionals, to the rehabilitation organizations, the insurance companies, to the government.    


\begin{figure}
  \includegraphics[width=\textwidth]{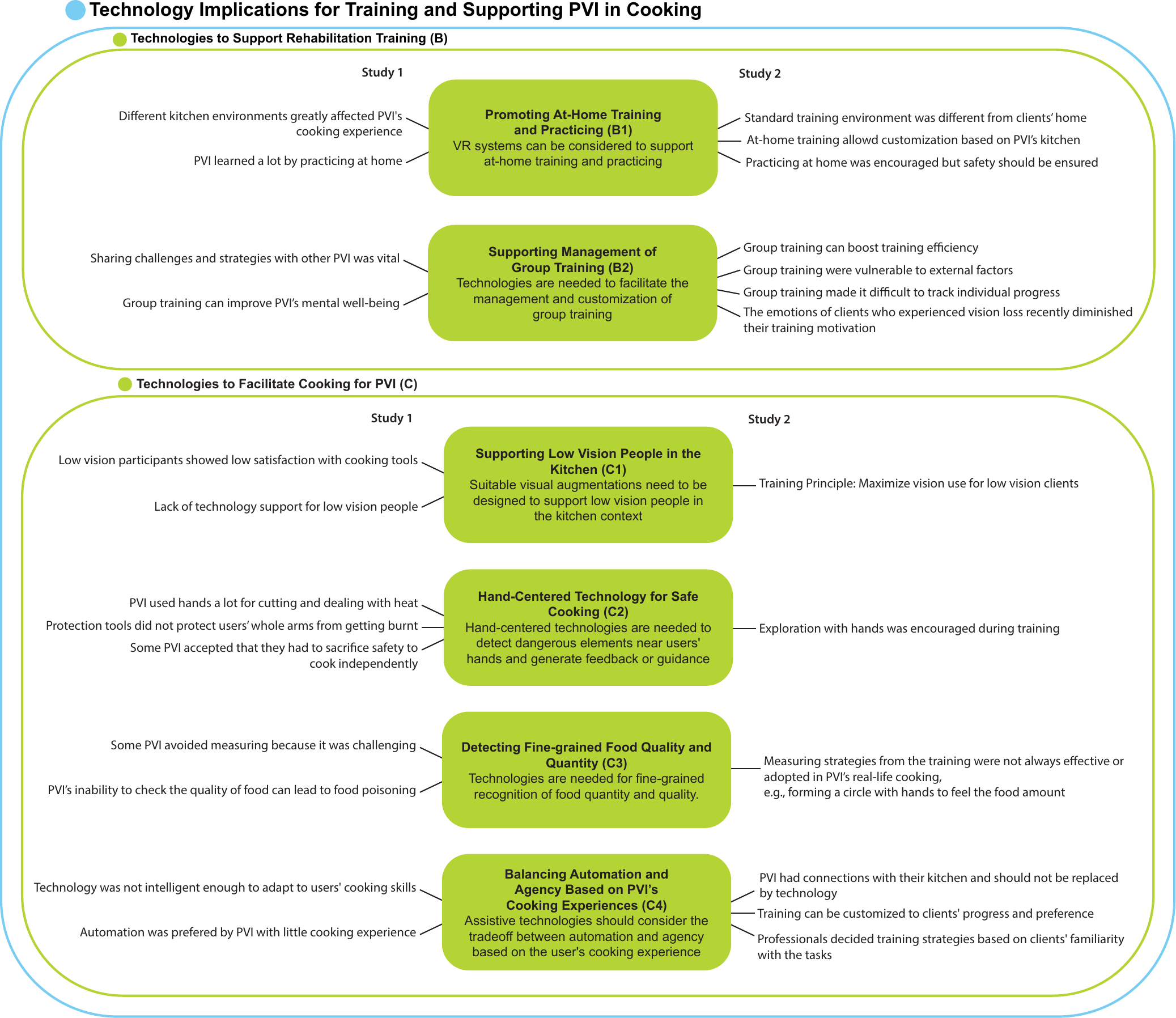}
  \caption{Technology implications to support rehabilitation training (B1-B3) and improve PVI's everyday cooking experiences (C1-C3).}
  \label{fig:connection_BC}
    \vspace{-3ex}

\end{figure}

\subsection{Technology Implications for Training and Supporting PVI in Cooking.}
Combining the professionals' suggestions and PVI's needs, 
we discuss the design considerations for technologies that support more effective and practical rehabilitation training (Section \ref{sec:tech4training}) as well as technologies that directly facilitate PVI's cooking process (Section \ref{sec:tech4cooking}). Figure \ref{fig:connection_BC} details the implications triangulated by the two studies. 


\subsubsection{Technologies to Support Rehabilitation Training}
\label{sec:tech4training}
We first discuss the potential technologies that can improve PVI's training experiences. 

\textbf{\textit{Promoting At-Home Training and Practicing (B1).}} To close the gap between the well-controlled training environment and PVI's own kitchen, 
at-home training and practicing could be a good solution. Via at-home training, professionals can customize the training plan and strategies based on the clients' home environment to maximize the training outcomes.
Moreover, at-home practicing could help PVI better apply their training knowledge in daily cooking.
However, at-home training is less affordable than standard training since it requires more effort from professionals, while at-home practicing could pose safety risks without proper supervision.

As such, future technologies should consider how to facilitate a safe and affordable at-home training and practicing for PVI. For example, virtual reality (VR) simulation could be a potential direction as VR devices become increasingly ubiquitous and affordable. Researchers have designed and built VR systems to support rehabilitation \cite{howard2017meta, schultheis2001application}, training \cite{cha2012virtual, xie2021review}, and remote mentoring and collaboration \cite{chen2022investigating, gasques2021artemis, radu2021virtual}, such as a VR simulator that supports ankle rehabilitation for patients who had stroke \cite{laver2017virtual}.
Such technologies have the potential to be adapted to the cooking training context for PVI to promote at-home training and practicing. For instance, different kitchen environments can be simulated to help PVI adapt to complex real-life kitchen situations more quickly, as well as ensuring their safety during practicing. VR or Mixed Reality training may also allow the professionals to connect with and mentor the clients remotely, thus reducing the time and commute costs. However, given the general accessibility issues of VR for PVI \cite{zhao2018enabling, mott2019accessible}, more research is needed to explore how to design an accessible and easy-to-setup VR rehabilitation system for PVI.


\textbf{\textit{Supporting Management of Group Training (B2).}}
Compared to one-on-one training, group training provides PVI opportunities to communicate and learn from others (Study 1) and increases the training efficiency by training multiple clients at once (Study 2). Prior work~\cite{alma2013effectiveness} has also found that group rehabilitation training can potentially improve the mental well-being of PVI, which is especially favorable for PVI who experienced vision loss recently. 
However, according to the rehabilitation professionals, such a training format is hard to manage given the different needs and learning abilities of different clients. Moreover, the group training is more susceptible to external factors, for example, many group sessions have been suspended due to the COVID-19 pandemic. Thus, it is important to consider how to facilitate the management and customization of the training in a group setting. 
For example, collaborative technologies could help PVI and professionals schedule and coordinate group sessions. Personalized devices (e.g., smartwatch, smartphone) could also be used as a complementary tool in the group training session to help monitor the progress of each client and keep the professionals updated. Systems that fulfill such purposes have been explored in the field of educational technology \cite{silva2014glassist, holstein2019co}. There is a potential to apply and adapt these technologies to support the group training sessions for PVI. 
Moreover, the aforementioned VR-based training can also be expanded to a group setting to remove the barriers brought by external factors such as pandemic and transportation.

\subsubsection{Technologies to Facilitate Cooking for PVI}
\label{sec:tech4cooking}
In addition to training and learning, more effective technologies are needed to directly support PVI in cooking. Via the two studies, we triangulated PVI's needs and distilled design implications for assistive technologies in the kitchen. Beyond confirming some similar insights to prior research \cite{li2021non, guo2017facade}, such as supporting hands-free interactions, making kitchen appliances accessible, and handling kitchen incidents, our work uncovered new technology implications based on PVI and professionals' feedback. We detail the unique implications below.

\textbf{\textit{Supporting Low Vision People in the Kitchen (C1).}}
Our research uncovered the major differences between blind and low vision people in cooking activities. Our research showed that low vision participants were significantly less confident than blind people in the kitchen. While no statistical significance, our participant sample also demonstrated a trend that low vision people felt less safe, more physically tired, and more mentally stressed than blind people when cooking.
Instead of relying on audio and tactile cues, low vision people looked for and even purposely created visual cues (e.g., color contrast, size contrast, reflections) to leverage their residual vision in the kitchen. 
The rehabilitation professionals in Study 2 also distinguished blind and low vision clients when planning training content, encouraging low vision clients to make the most of their residual vision in cooking. However, there is a lack of technology support for low vision people in the kitchen, resulting in their low satisfaction with cooking tools. 

While prior research has explored the challenges and needs of blind people during cooking \cite{li2021non} and designed technology for them \cite{guo2016vizlens, guo2017facade, sato2017navcog3}, little research had focused on the unique needs of low vision people. Recently, low vision research has attracted more attention with researchers exploring visual augmentations for low vision people in different tasks, including visual search and navigation \cite{zhao2016cuesee, zhao2020effectiveness}. However, no research has looked into cooking tasks. It is unknown whether existing visual augmentations could be applied to the crowded and busy kitchen context. Future research should consider low vision people's needs and design effective visual augmentations that fit for the kitchen context, such as, visual highlights that augment the position and contour of sharp tools, or color-coded augmentations to visualize elements with high temperature. 

\textbf{\textit{Hand-centered Technology for Safe Cooking (C2).}}
Kitchen safety is a critical issue emphasized by both PVI and rehabilitation professionals. The biggest safety concerns reported in both studies were cutting and interacting with heat, echoing prior research \cite{kashyap2020behaviors,li2021non}. Participants reported several cases of getting injured in the kitchen, such as cutting themselves when using a grater (David) or burning their hands when moving the pot on the burner (Judy). Importantly, we found that hands are at higher risks than other body parts since they are directly involved in most cooking tasks. Despite the potential safety risks, touching was encouraged by the rehabilitation professionals (e.g., Maria) since it was the easiest and most effective way to perceive the environment and complete cooking tasks. Thus, researchers should consider hand-centered technology to protect PVI's safety and guide them through the cooking process. 

To achieve this goal, artificial intelligence (AI) technologies should be involved to interpret the environment and recognize risks in the kitchen. Various computer vision and other sensing technologies have been explored for the kitchen scenario. For example, Zoran et al. ~\cite{zoran2015cutting} created a kitchen knife with sensing capacity to predict when fingers could be at high risk of injury; Green and Goubran~\cite{green2013thermal} and Maureillo et al. \cite{mauriello2019thermporal} used thermal cameras to detect heat sources (e.g., burners); datasets for cooking (e.g., the Epic Kitchens dataset) were also created to support recognition in the kitchen \cite{damen2020epic}. Along with the state-of-the-art hand recognition and tracking \cite{zhang2020mediapipe, cheng2022efficient}, these technologies and resources could be used to detect the sharp and hot kitchen elements near the PVI's hand and generate real-time feedback. For example, a wearable assistive device \cite{tapu2020wearable} (e.g., smartglasses \cite{zhao2020effectiveness, zhao2019designing}, smartwatch \cite{grussenmeyer2016feasibility}) can be designed to recognize the distance and relative position of the dangerous elements to the PVI's hand and guide the user to avoid the risks.

\textbf{\textit{Detecting Fine-grained Food Quantity and Quality (C3).}} Measuring and checking food quantity and quality are major challenges posed to PVI. Without effective support, many PVI choose to avoid these tasks during cooking and even lower the expectation on their food. 
While the state-of-the-art assistive technologies (e.g., OrCam~\cite{OrCam} and SeeingAI~\cite{SeeingAI}) support a high-level text recognition and object detection, they cannot address such fine-grained quantity and quality checking tasks. Beyond telling what food is in front of them, PVI need to know more details about their food, such as how much amount of ingredients they have put in a dish, whether their food is covered with mold, and even whether their dish looks aesthetically appealing. More technologies are needed to provide PVI such detailed information about their food. Fortunately, researchers in computer vision have already put efforts in food inspection~\cite{zhou2019application, brosnan2004improving}, such as detecting the freshness of food \cite{mery2013automated}, as well as food weight and amount estimation \cite{nyalala2021weight, ciocca2015food}. 
HCI and accessibility researchers should consider how to combine these technologies with suitable feedback and instructions, thus enabling PVI to make safe, tasty, and good-looking food.

\textbf{\textit{Balancing Automation and Agency Based on PVI's Cooking Experiences (C4).}}
Our findings revealed the tension between automation and agency in technology design, showing that PVI's preference between automation and agency highly depends on their cooking experiences. On the one hand, PVI with little cooking experience (e.g., Hilary, Derek) indicate their desire for a smart kitchen that completes cooking tasks for them automatically. On the other hand, PVI with rich cooking experiences believe that advanced technology should not entirely replace them since they have emotional connections with the kitchen and want to cook ``for the joy of it'' (Maria). With more and more intelligent cooking systems designed to guide sighted users \cite{sato2014mimicook} and PVI \cite{kim2022vision} through the cooking process (e.g., following a recipe, using utensils), it is important to explore how to balance the automation and agency for PVI in an intelligent system.
Inspired by the rehabilitation training where professionals adjust their strategies according to PVI's cooking experiences and abilities, researchers should consider ``people-aware'' assistive technology to determine \textit{what} assistance and \textit{how much} assistance to provide based on the user's personal needs, thus enabling PVI to accomplish kitchen tasks independently and allowing them to enjoy cooking.


\subsection{Limitations and Future Directions}
Our studies had limitations. First, due to the pandemic, we only observed 10 PVI and some were observed remotely via a video call, which could prevent us from observing some cooking details. Besides, most participants had rich cooking experience (only two were beginners), which may limit our data from cooking beginners who are visually impaired. In the future, we will recruit more participants with more diverse cooking experiences and visual abilities, and conduct all in-person studies to thoroughly observe participants during cooking. 
Second, in Study 1, we encouraged participants to think aloud during cooking so that researchers could better interpret participants' cooking behaviors. However, this may increase participants' cognitive load and potentially alter participants' behaviors. Future research should adopt other observation methods to reduce interruption to participants' cooking workflow and audience effects.
Third, our exploration of rehabilitation training mainly focused on the professional's perspective. In the next step, we will focus on the visually impaired clients' perspectives and fully understand their training experiences and potential barriers. Lastly, the current studies focused on participants in USA. However, people in other culture may have different food and cooking habits, thus facing different challenges and adopting different strategies \cite{kashyap2020behaviors}. Future research should further explore the experiences of PVI in different geographical environments, providing richer and more comprehensive data.

\color{black}
\section{Conclusion}
In this paper, we explored PVI's challenges and strategies in the kitchen and the technology needs to support cooking activities through two studies. First, we conducted a contextual inquiry study where we observed 10 PVI cooking dishes in their own kitchens and interviewed them about their cooking experience. We then conducted semi-structured interviews with rehabilitation professionals on the training process for cooking skills as well as their technology recommendations for PVI. From PVI's perspective, we revealed the major challenges and needs of both blind and low vision people in the kitchen context. From the professionals' perspective, we identified the current training practices and the gaps between training and reality. In the end, our work suggested design considerations for future assistive technologies for PVI in the kitchen.


\bibliographystyle{ACM-Reference-Format}
\bibliography{sample-base}

\appendix
\renewcommand\thefigure{\thesection.\arabic{figure}}    
\setcounter{figure}{0}  
\renewcommand\thetable{\thesection.\arabic{table}}    
\setcounter{table}{0}  
\section{Appendix}


\begin{table}[H]
\centering
\begin{tabular}{ll}
  \toprule
  \textbf{Category} & \textbf{Activities}\\
  \midrule
  Environment & Organizing and locating items \\
              & Navigating \\
              & Cleaning \\
\addlinespace \midrule
Preparation  & Cutting and chopping \\
             & Peeling \\
             & Measuring \\
             & Transferring \\
             & Spreading \\
             & Detecting food that gone bad\\
\addlinespace \midrule
Cooking     & Boiling\\
            & Frying\\
            & Baking\\
            & Turning foods \\
            & Testing food for doneness\\
\addlinespace \midrule
Cleaning                            & Surfaces\\
                                    & Dishes \\
  
  \addlinespace \midrule
Tools and appliances & Interacting with tools and appliances \\
\addlinespace \midrule
Reading & Accessing recipes \\
        & Reading package \\
\bottomrule
\end{tabular}
\caption{Kitchen tasks we covered in Study 1 Interview}
\label{tab:tasks}
\end{table}

  

\begin{center}
\begin{longtable}{lll}
\toprule
\textbf{Themes} &\textbf{Sub-themes}\\
\midrule
Training and learning &Learning cooking skills\\
&Cooking training \\
& Limited resources of training\\
\midrule\addlinespace
Kitchen navigation and exploration &Navigation \\
&Finding things\\
&Organizing\\
\midrule\addlinespace
Dealing with heat &Use of stovetop \\
&Use of oven \\
\midrule\addlinespace
Using sharp tools &Using knives \\
&Using other sharp tools: blender blade, peeler, grater \\

\midrule\addlinespace
Interacting with appliances & Difficulty with using appliance interfaces\\
&Inaccessible instruction \\
\midrule\addlinespace
Fine-grained ingredient manipulation & Transferring \\
 &Measuring\\
\midrule\addlinespace
Following recipes &Reading recipes \\
&Experimenting new recipes\\
\midrule\addlinespace
Food quality and satisfaction &Food safety \\
&Food satisfaction and aesthetics\\

\midrule\addlinespace
Kitchen hygiene &Washing dishes and cleaning countertops\\
&Cleaning knives \\
\midrule\addlinespace
Low vision people's experience in kitchen &Lighting \\
&Color contrast \\
&See small kitchen elements\\
\midrule\addlinespace
Technology in the kitchen & Smart technology in the kitchen \\
&Issues with current technology \\
&Desired Technology\\
\midrule\addlinespace
External Help &Food delivery service \\
&Home-care workers \\
&Family, neighbors, and friends \\
\bottomrule
\caption{Themes and Sub-themes generated from Study 1}\\
\label{tab:PVI_themes}
\end{longtable}
\end{center}

\begin{table}[h]
\centering
\begin{tabular}{ll}
  \toprule
  \textbf{Themes} & \textbf{Sub-themes}\\ 
  \midrule
  Training format   & One-on-one \\
                    & Small group \\
                    & Large group \\
\midrule\addlinespace
  Training prodcedure and timelines & Customized training plan\\
                                    & Standard training plan\\
\midrule\addlinespace
 Assessment methods & Pre-training\\
                    & During-training\\
                    & Post-training\\
\midrule\addlinespace
Training principles	& Ensure safety\\ 
                    & Maximize residual vision use\\
                    & Use other sensory channels\\
                    & Practice at home\\
                    & Involve other specialists\\
                    
\midrule\addlinespace 
Training instructions/strategies	            & Verbal \\
                    & Hands-on\\
                    & Tactile diagram \\
                    & Trial-and-error \\ 
\midrule\addlinespace
Training content    & Cutting skills\\
                    & Measuring skills\\
                    & Pouring/transferring skills\\
                    & Dealing with heat skills\\
                    & Checking food doneness skills\\
                    & Labeling skills\\
                    & Reading skills\\
\midrule\addlinespace 
Technology	        & Low technology recommendations\\
                    & Advanced technology\\
                    & Desired technology features\\
\bottomrule
\end{tabular}
\caption{Themes and Sub-themes generated from Study 2}
\label{tab:OT_themes}
\end{table}

\end{document}